\documentclass[11pt,twoside]{article} 
\usepackage{cspm-asp2006}
\usepackage{epsfig,graphicx,natbib,url}  
\usepackage{lscape} 
\begin{document}
\setcounter{page}{27}

\markboth{Rutten}{Observing the Solar Chromosphere}
\title{Observing the Solar Chromosphere}
\author{Robert J. Rutten} 
\affil{Sterrekundig Instituut, Utrecht University, The Netherlands\\
       Institutt for Teoretisk Astrofysikk, University of Oslo, Norway}

\begin{abstract} 
  This review is split into two parts: one on chromospheric line
  formation in answer to the frequent question ``where is my line
  formed'', and one presenting state-of-the-art imagery of the
  chromosphere.  In the first part I specifically treat the formation
  of the Na\,D lines, Ca\,II\,H\&K, and H$\alpha$.  In the second I show
  DOT, IBIS, VAULT, and TRACE images as evidence that the chromosphere
  consists of fibrils of intrinsically different types.  The
  straight-up ones are hottest.  The slanted ones are filled by shocks
  and likely possess thin transition sheaths to coronal plasma.  The
  ones hovering horizontally over ``clapotispheric'' cell interiors
  outline magnetic canopies and are buffeted by shocks, most violently
  in the quietest regions.

  In the absence of integral-field ultraviolet spectrometry, H$\alpha$
  remains the principal chromosphere diagnostic.  The required
  fast-cadence profile-sampling imaging is an important quest for new
  telescope technology.
\end{abstract}

\section{Introduction}
An excellent review of observational chromospheric issues and research
is given by Philip G. Judge in the 2005 NSO/Sacramento Peak workshop
proceedings
 (\cite{rjr-JudgeSPO2005};
 \url{http://download.hao.ucar.edu/pub/judge/judge_ws23.pdf}).
His major conclusion is that while chromospheric magnetism smoothes
out with height, its thermal fine structuring remains tremendous.
Here, I tread less wide by concentrating on diagnostics of this fine
structure, in particular those in the visible.  Long-duration
observing with Hinode's SOT and higher-resolution observing exploiting
adaptive optics at existing telescopes (DST, VTT, SST) and hopefully
at upcoming telescopes (GREGOR, NST, EST, ATST) employing Ca\,II\,H\&K,
the Ca\,II infrared lines, and above all the H\,I Balmer lines are
likely to provide new high-resolution vistas and understanding of the
chromosphere the coming years while, unfortunately, ultraviolet
spectrometry lags behind.  I split the discussion into a didactic part
on chromospheric line formation and a morphological part on
chromospheric scenes, concluded by a list of speculations and a brief
discussion of research approaches.

\begin{figure}
  \includegraphics[width=\textwidth]{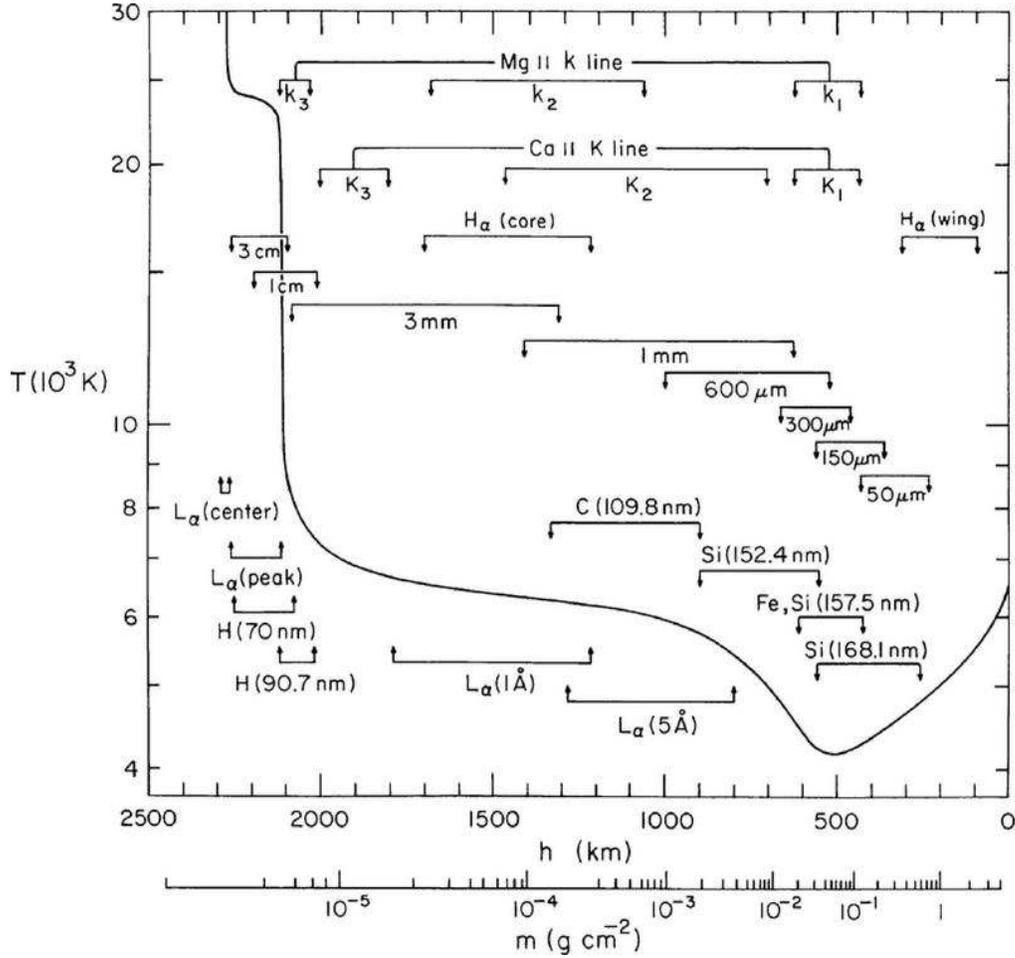}
  \caption[]{\label{rutten-fig:VAL3C}
  Classic VAL3C height-of-formation graph from 
  \citet{rjr-1981ApJS...45..635V}. 
  The bars span 90\% of the area under the intensity contribution
  functions ${\rm d} I/{\rm d} h$ plotted for many wavelengths in Fig.~36 of
  the same paper.
}
\end{figure}

\section{Chromospheric Line Formation} \label{rutten-sec:lineformation}
As author of lecture notes on radiative transfer 
  (\cite{rjr-Rutten2003:RTSA})
I am often asked how high some line is formed.  In
this section I follow the inspiring ``Fads and Fallacies'' example of
  \citet{rjr-Athay1976} 
to illustrate that the question is only answerable for numerical
models or simulations, not for observations.  

The answer evolved over the years from ``weighting functions'' 
  (e.g., \cite{rjr-1951AnAp...14..115P}, 
   \cite{rjr-Unsold1955})
through ``line-depression contribution functions'' 
  (e.g., \cite{rjr-1967AJ.....72R.793E}; 
   \cite{rjr-1974AZh....51.1032G}; 
   \cite{rjr-1986A&A...163..135M}; 
   \cite{rjr-1998A&A...332.1069K}) 
to ``response functions''
  (e.g., \cite{rjr-1971SoPh...20....3M}; 
   \cite{rjr-1975SoPh...43..289B}; 
   \cite{rjr-1977A&A....54..227C}) 
which became the backbone of least-square-fit inversions with the SIR
code of
  Ruiz-Cobo et al.\ (1990, 1992)
    \nocite{rjr-1990Ap&SS.170..113R} 
    \nocite{rjr-1992ApJ...398..375R}
and its companions
   (see
     \cite{rjr-2003AN....324..383D}; 
     \cite{rjr-2005A&A...439..687C}). 
I discuss the answer first in the context of solar-atmosphere model
evolution, and then for some specific lines.

\subsection{HOLMUL, SIR, VAL3C, RADYN, CO$^5$BOLD chromospheres}

For decades already, stellar abundance determiners prefer the HOLMUL
model of
  \citet{rjr-1974SoPh...39...19H} 
in the classical Uns\"old ``fine analysis'' recipe of computing
stellar line formation as plane-parallel hydrostatic LTE with best-fit
microturbulence and van der Waals damping enhancement but no other
parameters
  (cf.\  Rutten 1998, 2002).
     \nocite{rjr-Rutten1998a} 
     \nocite{rjr-2002JAD.....8....8R} 
The HOLMUL temperature stratification is essentially
  Holweger's (1967)
    \nocite{rjr-1967ZA.....65..365H}
empirical fit to many observed optical lines, from iron in
particular, assuming LTE excitation and ionization.  Actual
iron-line NLTE departures (mainly smaller opacity due to
overionization for weaker Fe\,I lines, source function deficit due to
scattering for the strongest Fe\,I lines, and source function excess
for pumped subordinate Fe\,II lines) happen to largely cancel in this
procedure over much of the optical spectrum
   (\cite{rjr-1982A&A...115..104R}), 
so that the model performs very well in reproducing any line similar
to the ones from which it was made, under the same assumptions.  It
does not possess a chromospheric temperature rise because iron lines
exhibit no self-reversals.

The Holweger technique of empirically establishing a
temperature-height stratification by fitting excitation temperatures
and optical depth scales to observed lines assuming LTE is automated
in the SIR code for Stokes evaluation.  It effectively constructs such
Holweger models per pixel.  They won't have chromospheres either,
except when inverting the Mg\,I 12-micron lines which do show
conspicuous reversals -- but then wrongly because their emission peaks
are actually photospheric NLTE ones
  (\cite{rjr-1992A&A...253..567C}).  
Another key inversion assumption is that solar stratifications vary
smoothly with height, as cubic splines through only a few sampling
nodes.  This is a dangerous assumption when large fluctuations occur,
such as the shocks making up umbral flashes
 (cf.\  \cite{rjr-2001ApJ...550.1102S}). 

The VAL3C model of
  \citet{rjr-1981ApJS...45..635V} 
differs from the Holweger approach in primarily fitting continua
rather than lines but extends much higher by including the full
ultraviolet, necessitating detailed NLTE evaluation of many bound-free
edges and PRD evaluation of Ly-$\alpha$.  This model is elevated to
stellar status (``the star VAL3C'') in my lecture notes because it
represents a complete self-consistent numerical simulation of the
radiation from a time-invariant plane-parallel star which strictly
obeys all standard equations in my course (plus some more, because the
latter do not yet treat PRD whereas the Avrett-Loeser PANDORA code
does) while resembling the sun (at least spectrally) in having a
chromosphere and some sort of transition region to a coronal regime.
The magnificent VAL3 paper adds a superb collection of informative
graphs diagnosing for many frequencies how this star radiates, where
its radiation originates, and breaking down its radiative energy
budget.  The FALC model of
  \citet{rjr-1993ApJ...406..319F} 
and more recent ones by
  e.g., \citet{rjr-2006ApJ...639..441F} 
represent updates of the approach with a similar code.
Figure~\ref{rutten-fig:VAL3C}, perhaps the best-known graph of solar
physics, shows heights of formation for many VAL3C diagnostics.  In
the star VAL3C these are rigorously correct.  Are they in the sun?

For example, the chromospheric dynamics analysis of 
  \citet{rjr-2004ApJ...604..936B} 
relied on the VAL3C height difference between H$\alpha$ and Ca\,II\,K
core formation, the latter higher than the former, to estimate upward
propagation speeds and so assign particular MHD wave modes and
appropriate mode conversions to a few observed oscillation wave trains
through wavelet analysis of network bright points in filtergrams.  The
approach is interesting but the reliance on VAL3C formation heights,
with Ca\,II\,K higher up than H$\alpha$, is na\"{\i}ve even if purely
radial structures were indeed observed in both lines (see
Fig.~\ref{rutten-fig:mosaic}).  I feel that numerical wave simulation
comparing detailed H$\alpha$ and Ca\,II\,K line synthesis using a
detailed model of such a structure is a sine-qua-non step in such mode
identification.

\begin{figure}
  \centering
  \includegraphics[width=13cm]{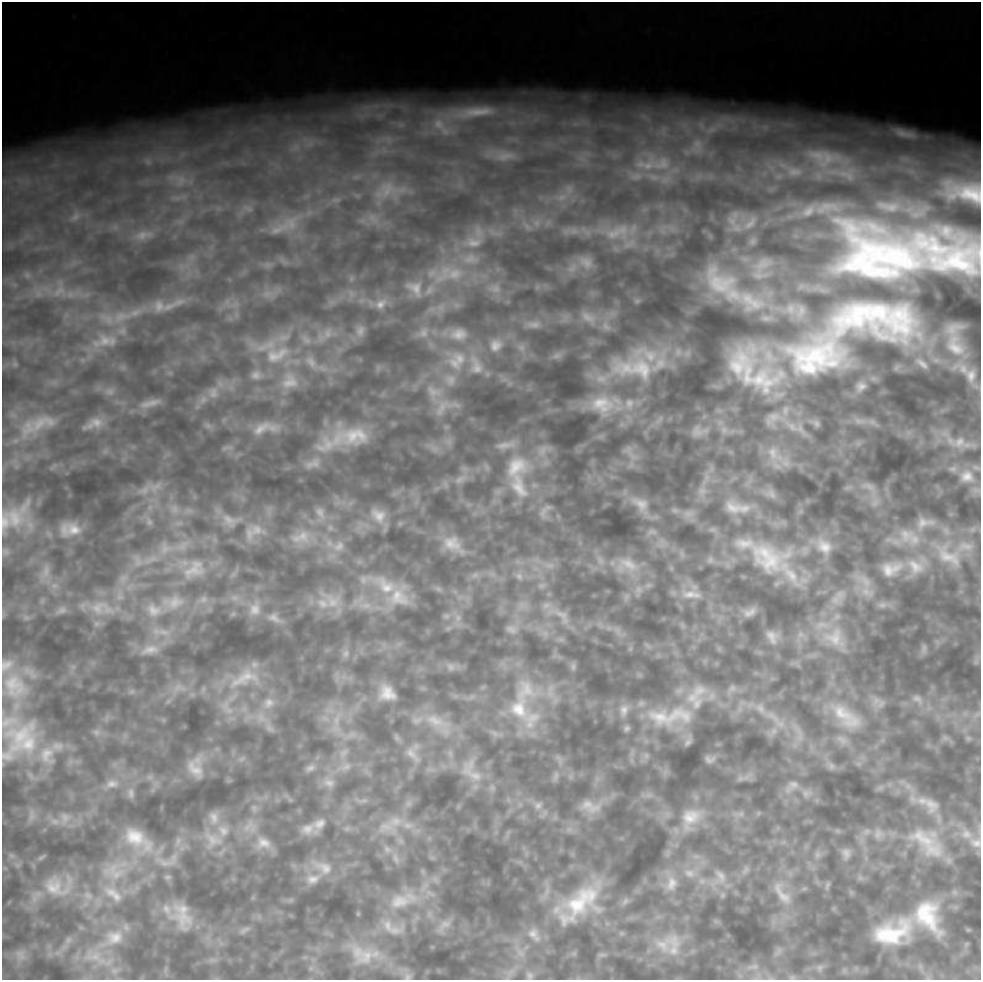}
  \caption[]{\label{rutten-fig:gillespie}
  Another classic: part of a Ca\,II\,K$_{2V}$ spectroheliogram taken by
  B.~Gillespie at Kitt Peak thirty years ago.  Other cutouts appeared
  on the cover of \citet{rjr-Lites1985a}, in \citet{rjr-Zirin1988}, in
  \citet{rjr-Rutten+Uitenbroek1991a}, as frontispiece to Solar Phys.~134,
  and elsewhere.  Outside active areas this image displays the solar
  clapotisphere rather than the solar chromosphere.  Courtesy
  K.\,P.~Reardon.
}\end{figure}

Detailed wave simulation was undertaken for the much easier case of
purely acoustic waves sampled by the 170\,nm and 160\,nm bandpasses of
TRACE by
  Fossum \& Carlsson (2005a, 2005b, 2006)
  \nocite{rjr-2005ApJ...625..556F} 
  \nocite{rjr-2005Natur.435..919F} 
  \nocite{rjr-2006ApJ...646..579F} 
using the RADYN code of 
  Carlsson \& Stein (e.g., 1992, 1997).
  \nocite{rjr-Carlsson+Stein1992b} 
  \nocite{rjr-1997ApJ...481..500C} 
These ultraviolet continua suffer considerable NLTE Si\,I bound-free
scattering, as is obvious in the pertinent Fig.~36 $B$-$J$-$S$ panels
of VAL3, but at least this scattering obeys complete frequency
redistribution over the ionization edge.  Fossum \& Carlsson included
it in evaluating brightness response to acoustic-wave perturbations.
They first explained the puzzling high-frequency phase-difference
behavior observed by
  \citet{rjr-2005A&A...430.1119D} 
as due to doubly-peaked 170\,nm response, and they then evaluated the
amount of observed high-frequency power as not enough to heat
the chromosphere.

The latter result is questioned on 
p.~93\,ff
of this book by
  \citet{rjr-Wedemeyer++2007} 
who compute synthetic 160\,nm images from CO$^5$BOLD simulations in
LTE, without and with magnetic fields, to claim that the RADYN code by
being only one-dimensional severely underestimates acoustic heating
through the small-scale high-frequency acoustic interference patterns
found in the 3D CO$^5$BOLD atmosphere at about $h\!=\!500$~km.
I have called such a quiet-sun regime where wave interference
acts as the dominant structuring agent, above the overshooting
convection but still below a magnetic canopy, a ``{\em clapotisphere\/}''
   (\cite{rjr-Rutten1995b}), 
inspired by 
  \citet{rjr-Carlsson1994} 
and the reference in 
  \citet{rjr-Rutten+Uitenbroek1991b} 
to 
   \citet{rjr-Dowd1981}. 
While the grey CO$^5$BOLD star does not (yet) possess a fibrilar
magnetism-dominated chromosphere as the solar one displayed in
Section~\ref{rutten-sec:scenes} below, it surely has a violent
clapotisphere, even in excess of RADYN's.

Does the sun possess a clapotisphere?  Yes, see the much-published
spectroheliogram in Fig.~\ref{rutten-fig:gillespie}.  Selecting the
special Ca\,II\,K$_{2V}$ passband emphasizes acoustic shocks as ``cell
grains'' in internetwork areas
  (\cite{rjr-Rutten+Uitenbroek1991b}). 
They are brightest in H$_{2V}$ and K$_{2V}$ through intricate vertical
shock interference explained beautifully in Figs.~4--7 of
  \citet{rjr-1997ApJ...481..500C}, 
but remain visible through wider passbands
  (Fig.~10 of \cite{rjr-2001A&A...379.1052K}). 

Which of the VAL3C, RADYN, and CO$^5$BOLD chromospheres resembles
the solar one the best?  In the VAL3C star the resonance lines of
Mg\,II and Ca\,II are the major chromospheric cooling agents and
require NLTE--PRD synthesis.  RADYN neglects h\&k and assumes CRD for
H\&K (perhaps mutually corrective since line cooling is too large in
CRD due to deeper photon escape through the wings) but adds the virtue
of computing time-dependent ionization.  CO$^5$BOLD has the virtue of
being 3D but unrealistically assumes grey LTE radiative transfer at
all heights, ignoring actual strong-line cooling, lack of surface
cooling through scattering, and fast-versus-slow asymmetry between
ionization and recombination in and behind shocks
  (\cite{rjr-2002ApJ...572..626C}). 
It may so overestimate the occurrence of fine structure
  (\cite{rjr-2006A&A...460..301L}). 

The verdict is not in.  Let us regard VAL3C, RADYN, and CO$^5$BOLD as
interesting stars with sun-like photospheres but chromospheres that
exist only computationally.  VAL3C assumes that temperature
fluctuations around the mean are small enough to be meaningfully
averageable.  RADYN's shocks deny such averaging but do not generate 
ultraviolet line emission.  CO$^5$BOLD has more acoustic heating but
its grey-LTE-instantaneous radiative cooling is very non-solar.

My feeling remains that the actual solar chromosphere is hot where the
field points up or fans out from network and plage, that the
internetwork is cool below the canopy except within shocks, that the
latter clapotispheric domain represents a wide radial gap in H\&K
emissivity and H$\alpha$ opacity, and that chromospheric internetwork
radiation in H$\alpha$, the Ca\,II infrared lines and all ultraviolet
lines including Ly-$\alpha$ comes from canopy-constituting
mottles/fibrils/spicules and their sheath-like boundary layers.  I
gave some evidence in
  \citet{rjr-Rutten2007a} 
and give some more in Section~\ref{rutten-sec:scenes} below.  

\begin{figure}
  \centering
    \includegraphics[width=\textwidth]{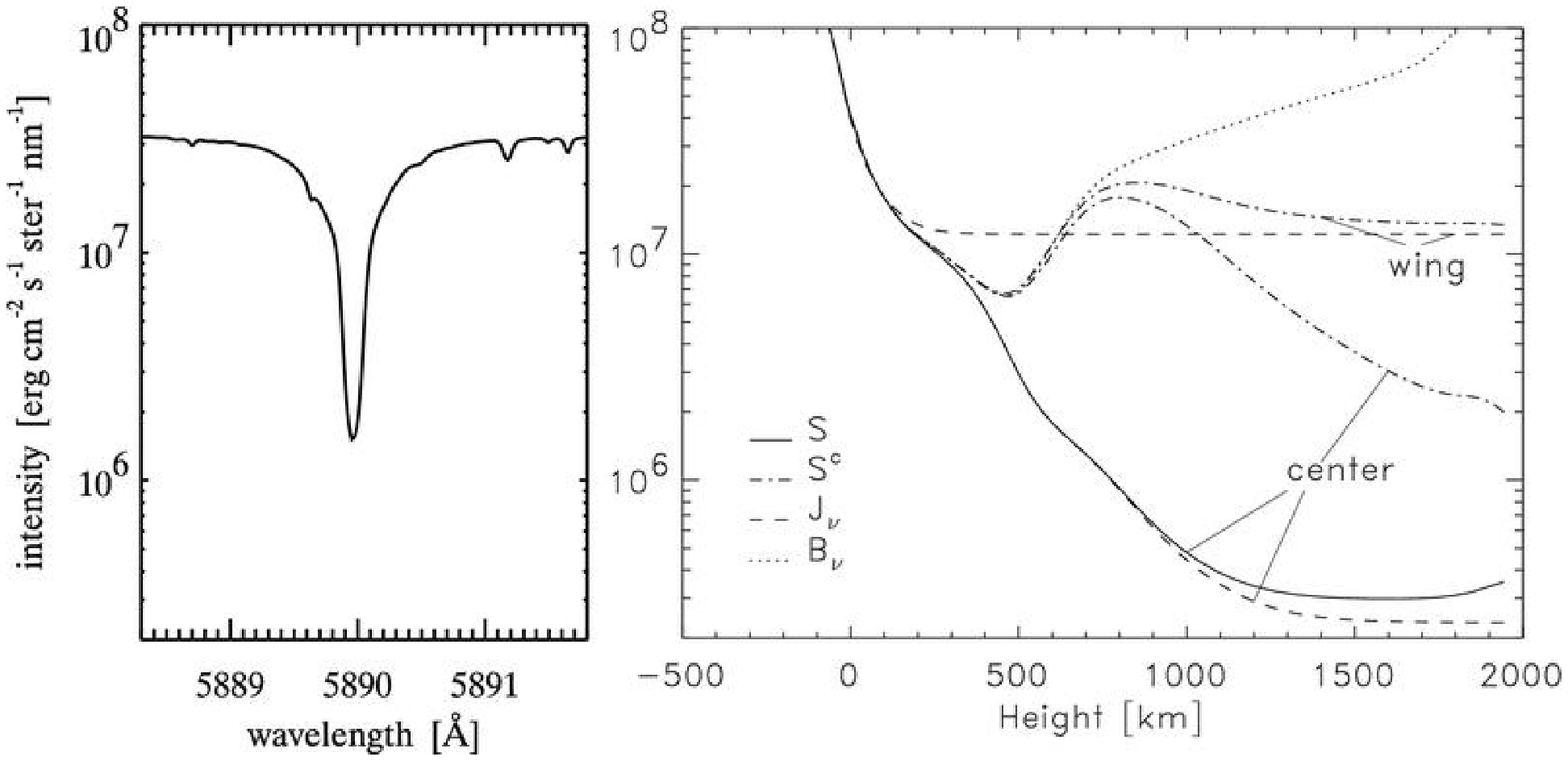}
  \caption[]{\label{rutten-fig:nad}
  Formation of the Na\,D$_2$ line.  Lefthand panel: disk-center line
  profile from the Kitt Peak FTS atlas
  (\cite{rjr-Neckel1999}) 
  plotted on the same logarithmic scale as the $S$,$B$,$J$
  line-formation graph at right from
  \citet{rjr-1992A&A...265..268U}. 
  The Na\,D$_2$ line is deep because its source function sinks deep,
  far below the temperature minimum.  The model shown by the dotted
  $B_\nu$ curve is actually FALC of
  \citet{rjr-1993ApJ...406..319F}, 
  whose photosphere copied the VAL3C update of
  \citet{rjr-1986ApJ...306..284M} 
  which effectively brought the significantly-cooler-than-HOLMUL upper
  photosphere of VAL3C back to HOLMUL (and to
  near-radiative-equilibrium and near-LTE iron ionization) through
  the inclusion of a great many NLTE-scattered ultraviolet line haze
  lines in PANDORA
 (cf.\ 
   \cite{rjr-Avrett1985}; 
   \cite{rjr-Rutten1990c}), 
  but Uitenbroek \& Bruls maintained the name VALC.  At line center,
  the total source function equals the line source function which
  closely mimics a two-level atom dominated by scattering with
  complete redistribution.  You may refresh your grasp of line
  formation by estimating $\varepsilon$ and working out why the
  background continuum source function differs between line center and
  wing (hint: why does $S_{\rm wing}^c \Rightarrow J_{\rm wing}$ at
  right?).  }
\end{figure}

\subsection{Na\,I\,D\ in VAL3C}

Figure~\ref{rutten-fig:nad} diagnoses Na\,I\,D formation in VAL3C
didactically.  The Eddington-Barbier relation $I_\nu \approx
S_\nu(\tau_\nu \!=\! 1)$ suggests, through simply drawing a horizontal
line from line center at left to the solid curve at right, that Na\,D$_2$
line center originates at $h \approx 600$~km.  It is more like $h
\approx 800$~km in
  \citet{rjr-1992A&A...265..237B} 
but in any case $\tau_\nu\!=\!1$ lies in the VAL3C chromosphere,
i.e.\ above the temperature minimum.  Contribution functions to the
emergent intensity or to the line depression will put the core
formation there too.  However, the intensity is obviously dominated by
scattered photons originating much deeper: the line-center source
function doesn't appreciate that VAL3C possesses a chromosphere.
The Na\,D lines are often called chromospheric but their VAL3C
brightness response is photospheric.  Only when studying Na\,D
Dopplershift may one call these lines chromospheric (assuming a
VAL3C-like chromosphere) because the Dopplershift is encoded at the
last photon scattering towards the observer.  Thus, the
``magnetoacoustic portals'' analysis of
  \citet{rjr-2006ApJ...648L.151J} 
comparing Na\,I to K\,I Dopplershifts rightfully claims measurement at
the bottom of the chromosphere, whatever that may be.
 
Let me ask you some examination questions.  Where is $\tau_\nu = 1$
for the weak blend at $\lambda = 5888.6$\,\AA\ in the lefthand panel of
Fig.~\ref{rutten-fig:nad}?  Should you compute an intensity or a line
depression contribution function for it?  Is Milne-Eddington or
Schuster-Schwarzschild the better approximation for inversion?  Well,
drawing the Eddington-Barbier connecting line locates $\tau_\nu=1$ at
about $h=100$~km, right?  Fitting a Milne-Eddington line-depression
contribution would be your first bet in modeling such a weak line,
right?  Wrong!  And not a little either.  The line is due to water
vapor in our own atmosphere, a million times higher up than your
estimate, and Schuster-Schwarzschild is the better description.

\begin{figure}
  \centering
  \includegraphics[width=\textwidth]{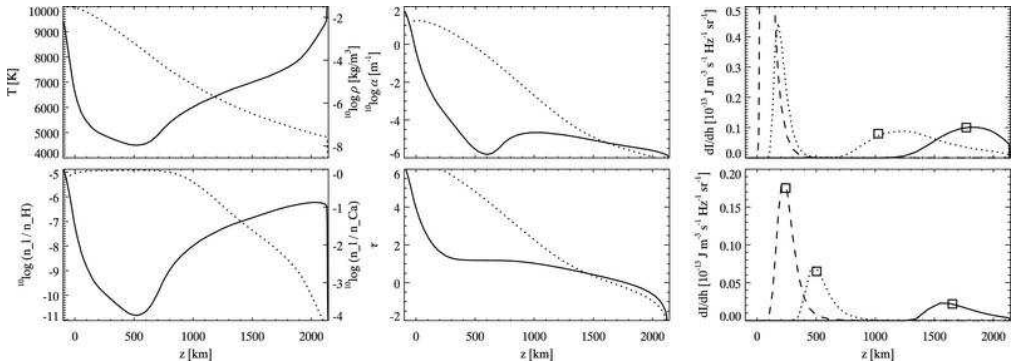}
  \caption[]{\label{rutten-fig:val-ha}
  Didactic explanation of the difference between H$\alpha$ and Ca\,II\,H
  formation assuming LTE, taken from
  \citet{rjr-2006A&A...449.1209L}. 
  {\em Top left\/}: FALC temperature (lefthand scale, solid) and
  density (righthand scale, dotted).  {\em Bottom left\/}: population
  fraction of the lower level relative to the total species density,
  respectively for H$\alpha$ (lefthand scale, solid) and for Ca\,II\,H
  (righthand scale, dotted).  {\em Top center\/}: line-center
  extinction coefficient for H$\alpha$ (solid) and Ca\,II\,H (dotted).
  {\em Bottom center\/}: line-center optical depth for H$\alpha$
  (solid) and Ca\,II\,H (dotted).  {\em Top right\/}: H$\alpha$ intensity
  contribution functions for $\Delta \lambda = 0$ (solid), $-0.038$
  (dotted), and $-0.084$\,nm (dashed) from line center.
  Squares mark $\tau \!=\! 1$ locations.  {\em Bottom
  right\/}: the same for Ca\,II\,H, at $\Delta \lambda \!=\! 0$ (solid),
  $-0.024$ (dotted), and $-0.116$\,nm (dashed) from line center.
}
\end{figure}

\subsection{Ca\,II\,H and H$\alpha$ in FALC}
Figure~\ref{rutten-fig:val-ha} compares H$\alpha$ formation to Ca\,II\,H
formation in FALC assuming LTE which holds reasonably well in the
wings of these lines.  The difference is enormous!

Ca\,II\,H originates from the ground state of the dominant ionization
stage, containing virtually all calcium particles out to $h\approx
800$~km where ionization (taking only 11.9~eV) to Ca\,III\ sets in (row
2 panel 1), and so has smooth sampling throughout the lower atmosphere
with near-constant inward $\tau_\nu$ build-up, resulting in nice
single-peaked intensity contribution functions which smoothly shift
outward closer to line center.  The extended Ca\,II\,H\&K wings present
an ideal probe to step smoothly through the lower atmosphere, also
furnishing blends for Doppler and Zeeman sampling as diagnostics ripe
for SIR inversions.

In contrast, H$\alpha$ originates from the extremely
temperature-sensitive $n\!=\!2$ level at 10.2~eV, and so has a
pronounced formation gap around the temperature minimum, no $\tau_\nu$
buildup there, and double-peaked contribution functions as already
shown by
  \citet{rjr-1972SoPh...22..344S}: 
in H$\alpha$ one either observes the deep photosphere or the overlying
chromosphere.

\section{Chromospheric Scenes} \label{rutten-sec:scenes}
\begin{figure}
  \centering
  \includegraphics[width=\textwidth]{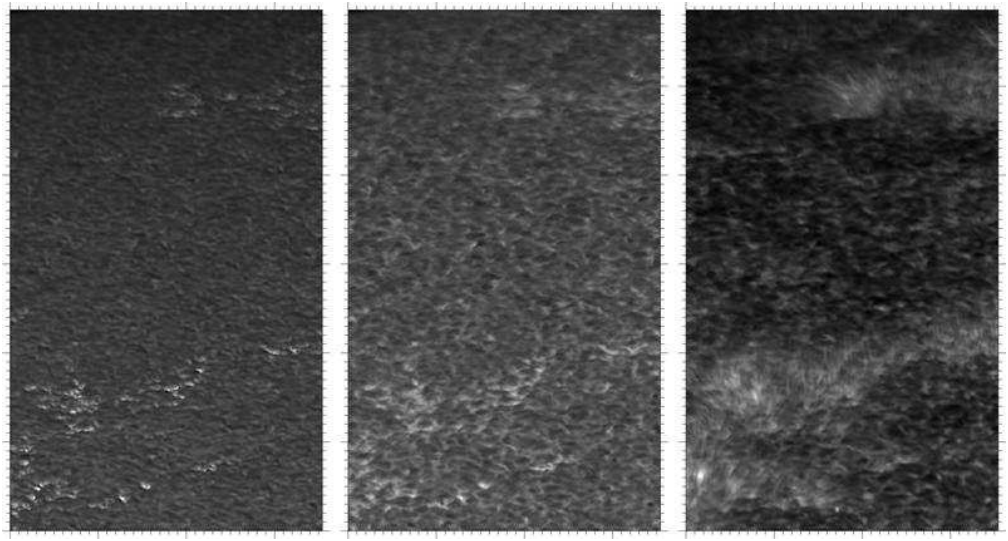}
  \caption[]{\label{rutten-fig:straws}
  Three partial near-limb images taken with the DOT on June 18, 2003.
  Ticks at arcsec intervals.  {\em Left\/}: G band.  {\em Center\/}:
  Ca\,II\,H wing.  {\em Right\/}: Ca\,II\,H line center.  From
  \citet{rjr-Rutten2007a}. 
}\end{figure}

\begin{figure}
  \centering
  \includegraphics[width=\textwidth]{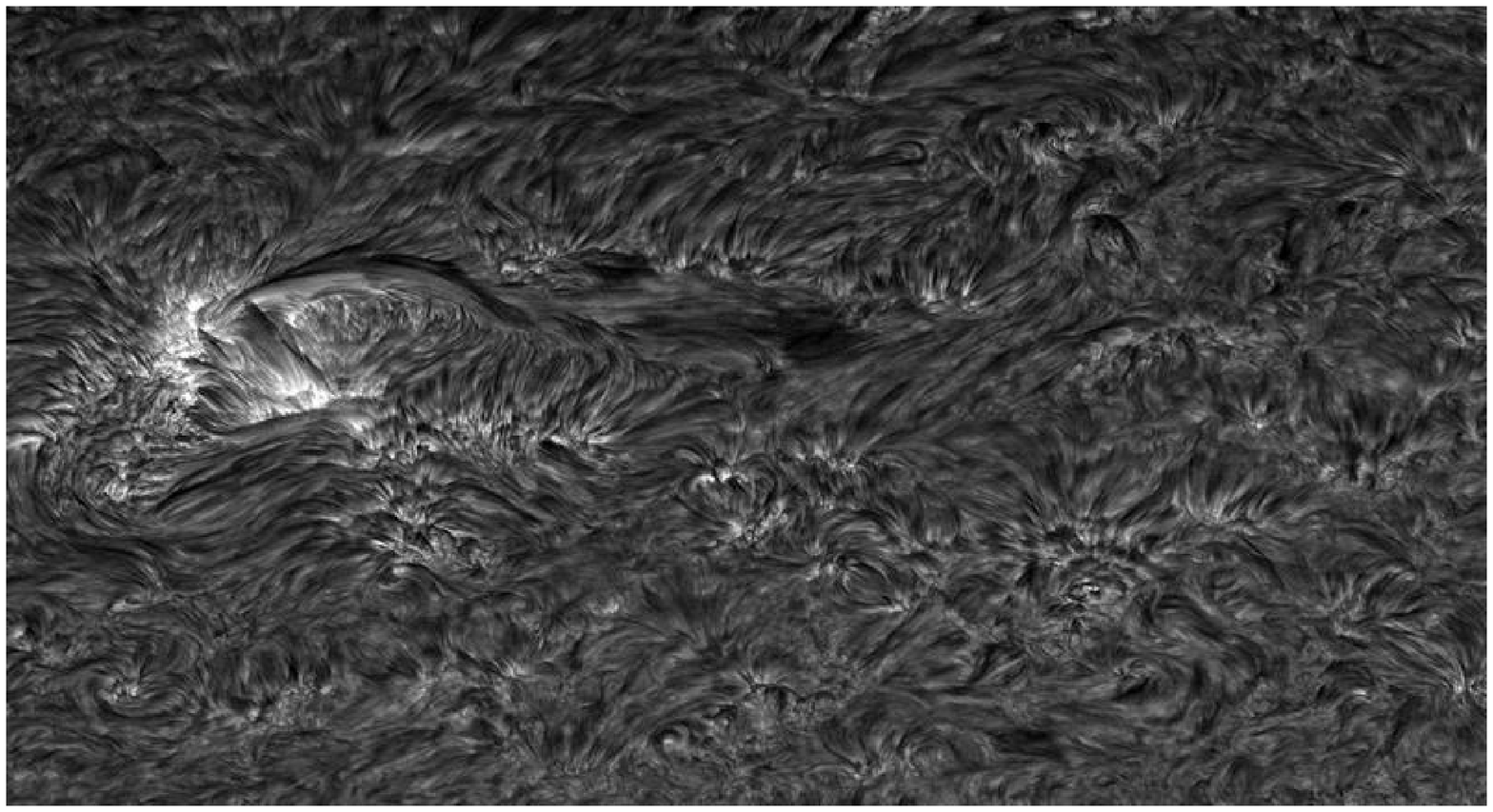}
  \vspace{2mm}
  \includegraphics[width=\textwidth]{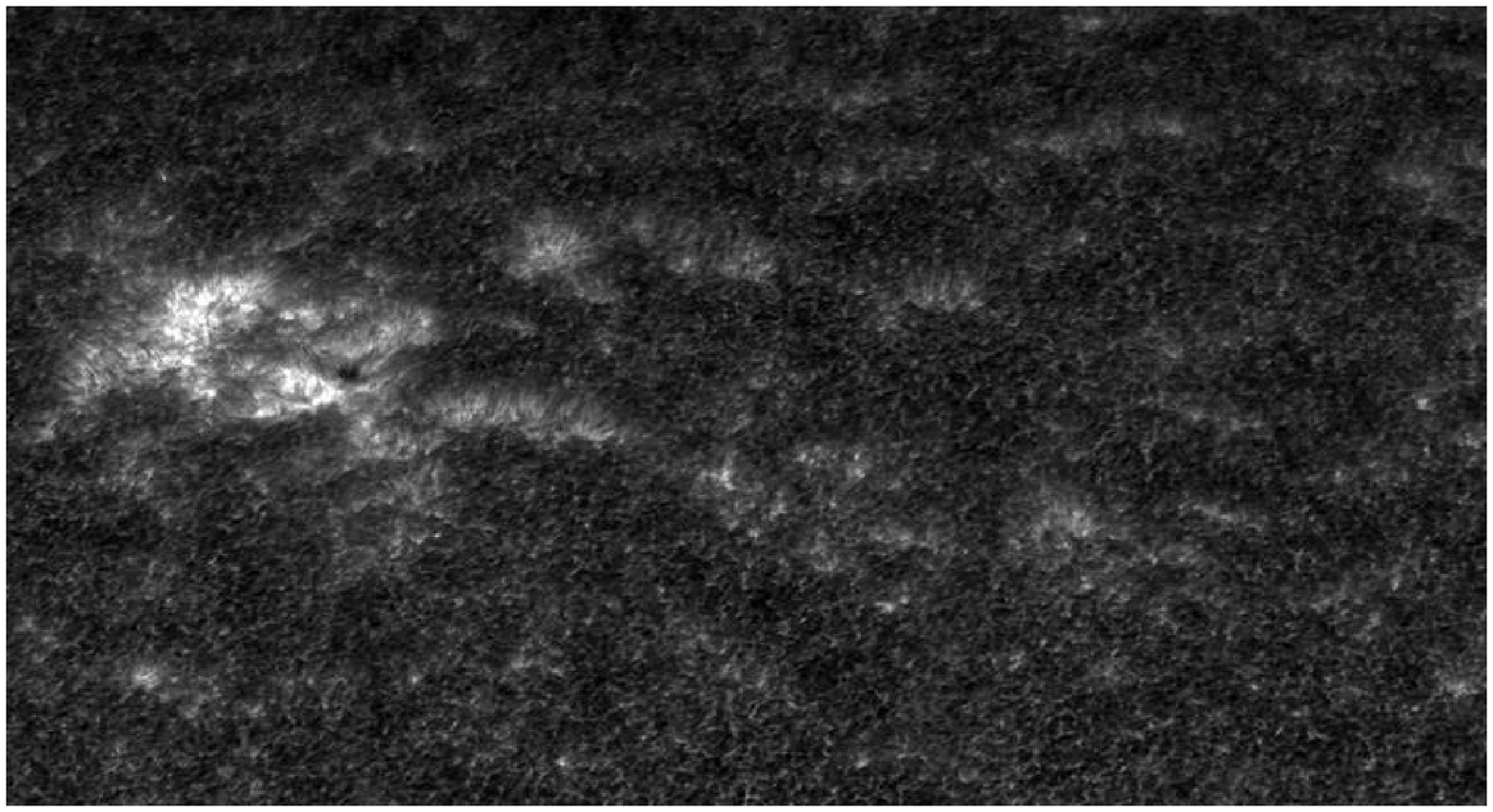}\\
  \caption[]{\label{rutten-fig:mosaic}
  Simultaneous image mosaics taken with the DOT on October 4, 2005,
  respectively in H$\alpha$ and Ca\,II\,H.  The field of view is close to
  the limb (off the top) and measures about $265\times 143$~arcsec$^2$.
}\end{figure}

\begin{figure}
  \centering
  \includegraphics[height=80mm]{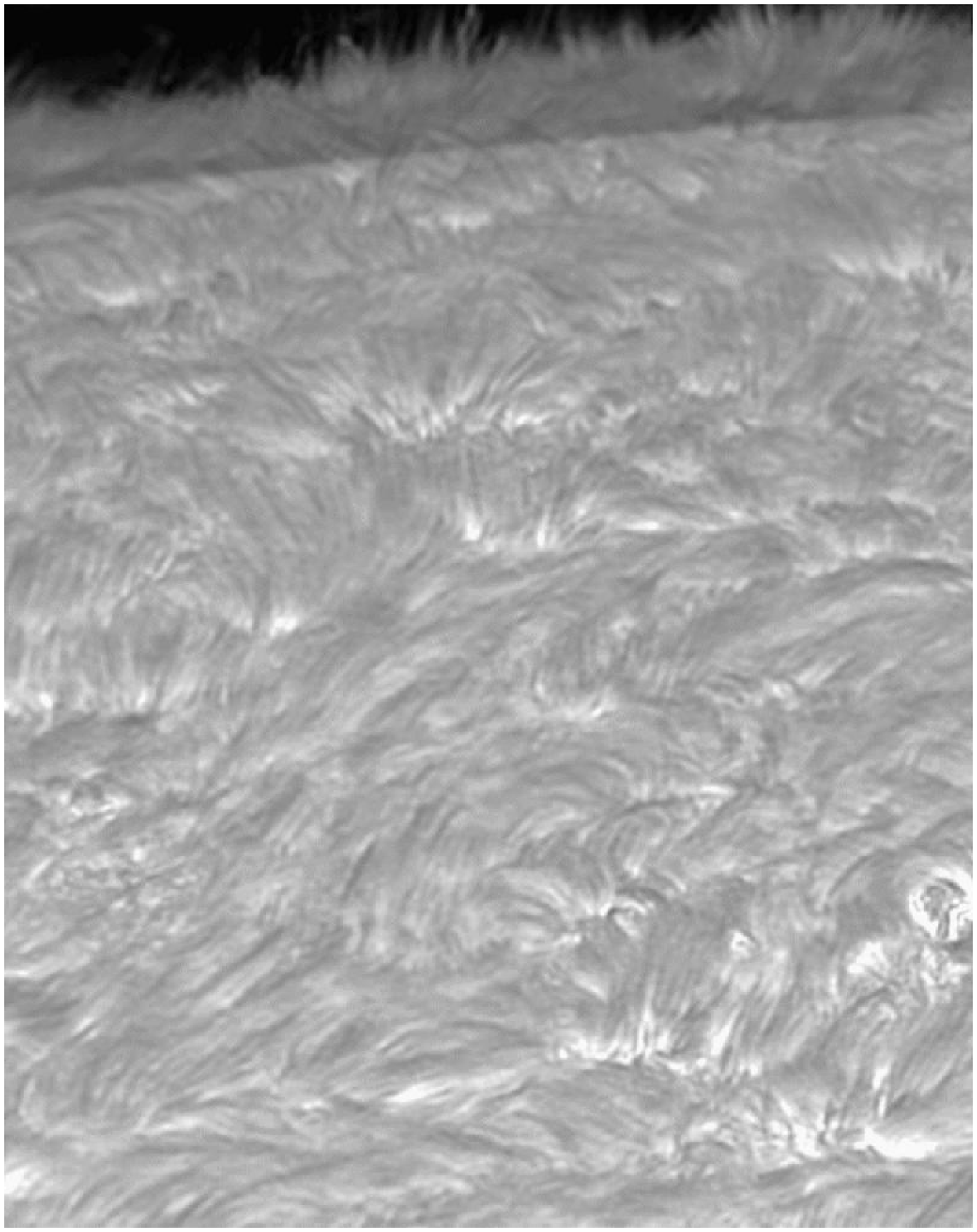}
  \hspace{4mm}
  \includegraphics[height=80mm]{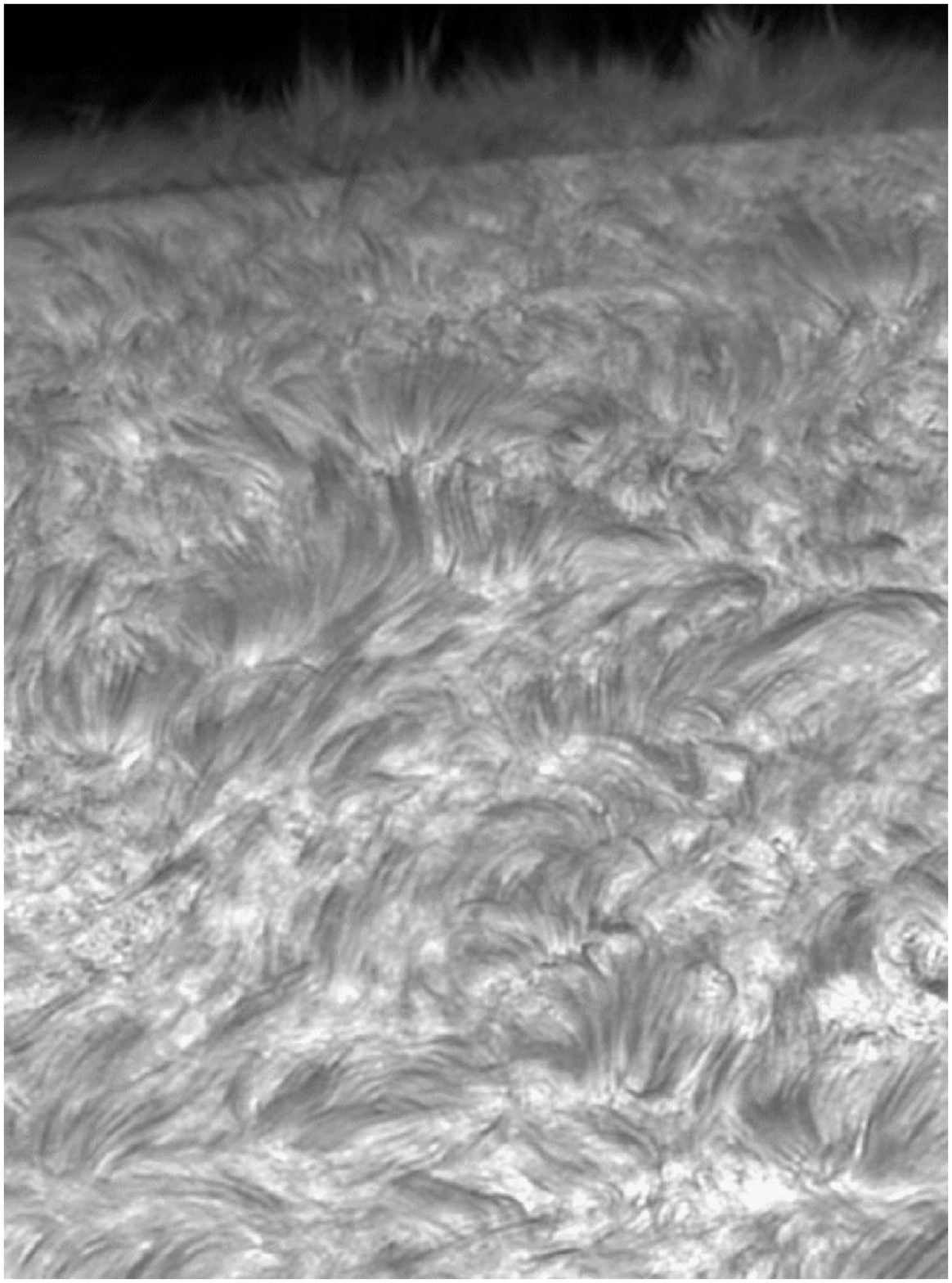}\\[2mm]
  \includegraphics[height=80mm]{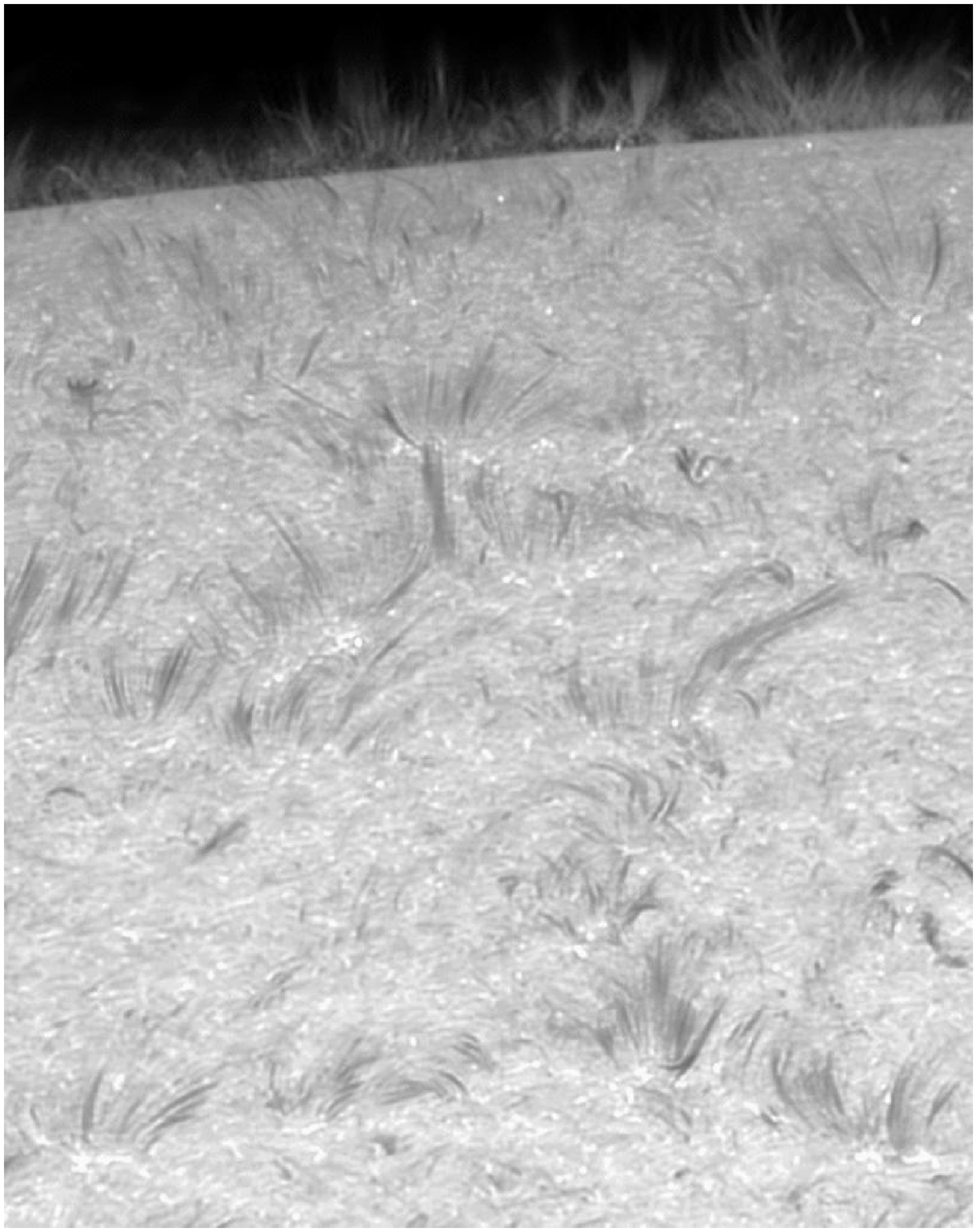}
  \hspace{0mm}
  \includegraphics[height=80mm]{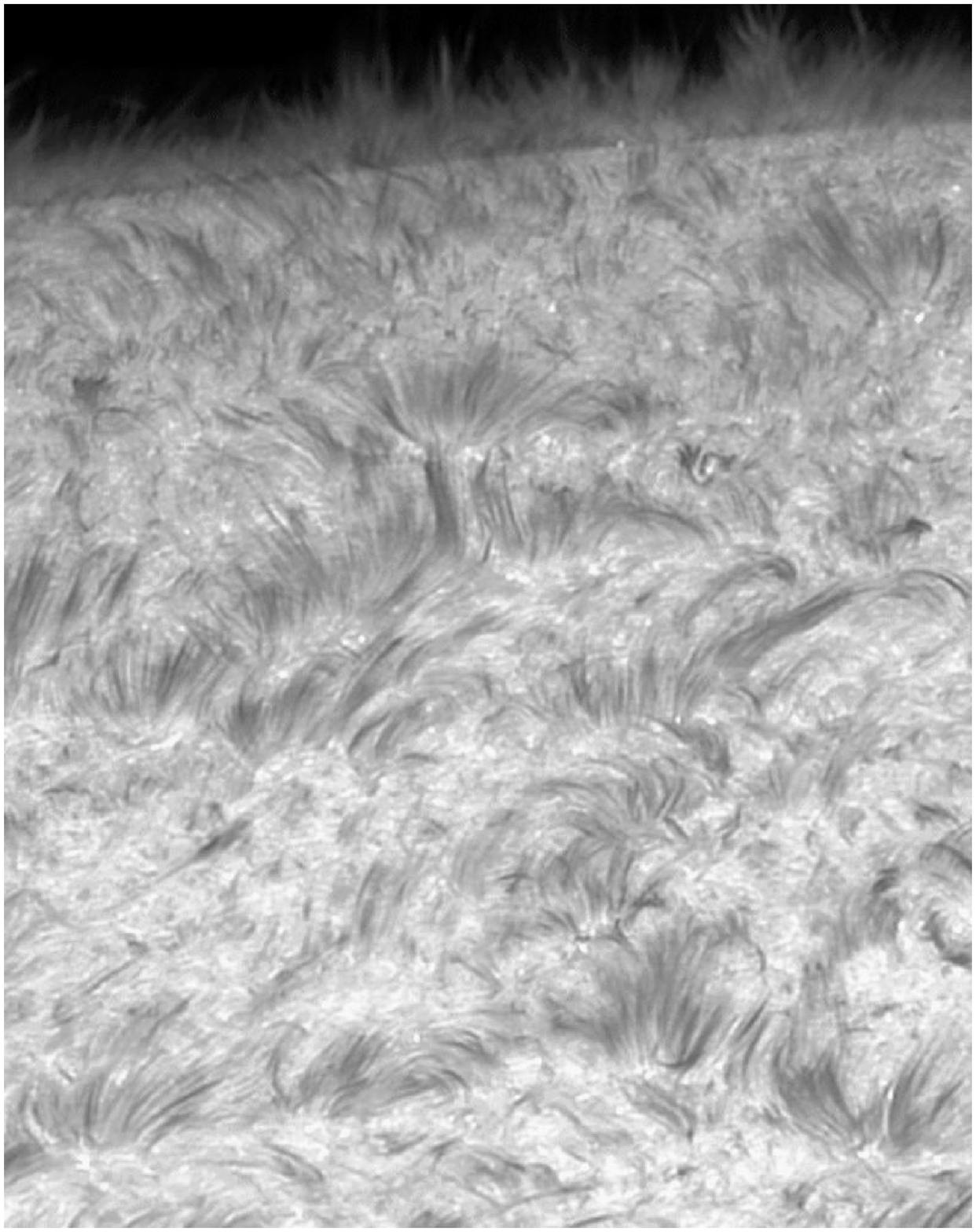}
  \caption[]{\label{rutten-fig:limb}
  Profile-sampled H$\alpha$ fine structure near and at the limb, taken
  with the DOT on October 4, 2005.  Clockwise: line center, $\Delta
  \lambda = -400$, $-600$, $-800$~m\AA, not simultaneous but all four
  taken within one minute.  Field of view about $70\times85$~arcsec$^2$.
}\end{figure}

In this section I insert images from the DOT\footnote{All 
  available at \url{ftp://dotdb.phys.uu.nl} with a convenient
  search interface at \url{http://dotdb.phys.uu.nl/DOT.}}, 
IBIS, and VAULT to demonstrate that the chromosphere is intrinsically
fibrilar (with the term ``fibrils'' encompassing quiet-sun ``mottles''
and off-limb ``spicules'').  This message is far from new
  (e.g., \cite{rjr-Zirin1988})
but the state-of-the-art images in
Figs.~\ref{rutten-fig:straws}--\ref{rutten-fig:vault} deliver it
beyond any wishful 1D thinking.

\subsection{Ca\,II\,H\&K and H$\alpha$ chromosphere}

Figure~\ref{rutten-fig:straws} shows at left the onset of reversed
granulation and ``faculae'', the latter as short bright stalks where
our slanted G-band view penetrates through relatively empty network fluxtubes
into hot granule tops as sketched in Fig.~7 of
    \citet{rjr-Rutten1999d} 
and more recently in Fig.~4 of
    \citet{rjr-2004ApJ...607L..59K}. 
The center panel shows a similar scene sampled slightly higher up.
The third panel shows a dark wing-contributed background of reversed
granulation with some shock interference wherever there is
insignificant magnetic-feature emission in the Ca\,II\,H core.  The
active network shows up through clusters of long, thin, bright
features added by the line core.  They start at photospheric faculae,
are sharply delineated from the dark background in their foreground,
appear to be optically thin, and stand rather upright causing overall
hedge-row appearance.  They make up the bright patches in the
narrower-band image in Fig.~\ref{rutten-fig:gillespie}.  The movie
from which this frame is taken shows that they are very dynamic.  I
called them ``straws'' in
  \citet{rjr-Rutten2007a}. 
They are only seen at high resolution, meaning angular rather than
spectral resolution.  In fact, it is better to use a fairly wide
passband to catch the emission peak whatever its Dopplershift.  Adding
inner-wing reversed-granulation background poses no problem because
this is quite dark.

Figure~\ref{rutten-fig:mosaic} is a similar but higher-up near-limb
scene comparison, between Ca\,II\,H and H$\alpha$.  The Ca\,II\,H image
again shows much dark photospheric background with bright network
hedges and straw crowding in the small active region.  In contrast,
H$\alpha$ is chromospheric nearly everywhere, with fibrils covering
internetwork cells in a fibrilar canopy.  The comparison immediately
repudiates the notion that Ca\,II\,H is formed higher than H$\alpha$.  The
lower panel shows only the onset of the chromosphere, the upper one
the full monty.

Figure~\ref{rutten-fig:limb} samples the limb part of this field while
stepping the DOT H$\alpha$ filter through the line.  The line-center
image again shows a mass of cell-spanning fibrils as a flattened
carpet, with upright ones jutting out from network.  I doubt that the
double limb is caused by parasitic light (continuum leak outside the
H$\alpha$ passband): the lower limb is bumpy (zoom in with a viewer)
and seems to mark the top of the carpet, the upper limb the end of
hedge-row visibility.  We have no DOT movie yet of off-limb spicules,
but many are bound to bounce up and down with 3--5~min periodicity --
perhaps fading on their return to convey the classical notion that
spicules send up much more mass then comes down.  Their upper ends
correspond to the tops of ``dynamic fibrils'' as those observed
on-disk with the SST on the same glorious day and analyzed in detail
in the beautiful paper of
  \citet{rjr-2007ApJ...655..624D}. 
%

\begin{figure}
  \centering
  \includegraphics[width=\textwidth]{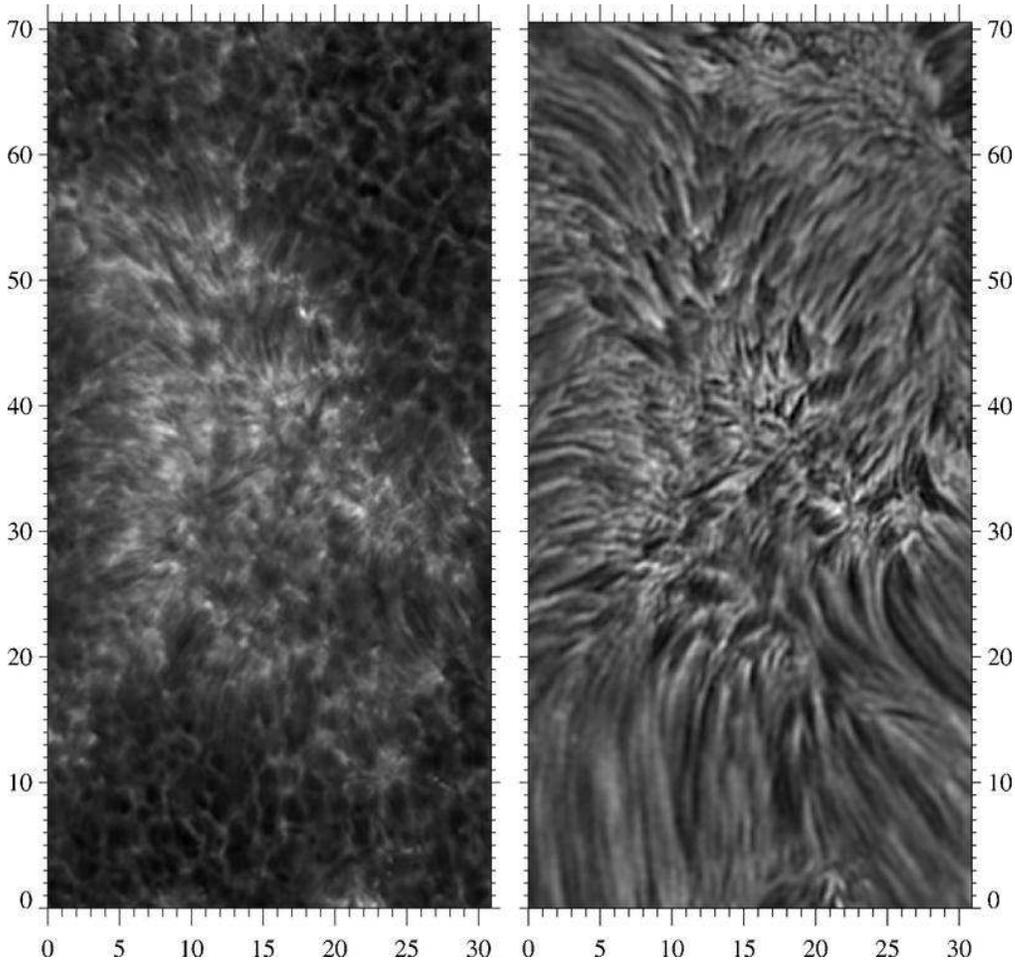}
  \caption[]{\label{rutten-fig:network}
  Two partial on-disk images taken with the DOT on April 24, 2006.
  Scales in arcsec.  {\em Left\/}: Ca\,II\,H. 
  {\em Right\/}: H$\alpha - 350$~m\AA.   Courtesy P.~S\"utterlin.
}\end{figure}

The off-limb line-center fibrils constitute the authentic
``chromosphere'' because this name comes from their H$\alpha$ emission
during totality.  Since there is no intrinsic difference between
off-limb fibrils and on-disk fibrils (they don't care about our
location in their sky), the proper definition of ``chromosphere'' is
simply the mass of fibrils observed in H$\alpha$.

The other panels show the same scene progressively further out into
the blue H$\alpha$ wing.  At decreasing line opacity one sees more and
more photospheric background (not even granularly reversed) between
the hedge rows.  In the outer wing only rather upright fibrils in
network hedge rows remain, appearing as dark to very dark strands
against the deep-photosphere background and across the continuum limb.
J.~Leenaarts has pointed out that the latter is much brighter than the
high-photosphere background seen in Ca\,II\,H, making straws dark in
H$\alpha$ while bright in Ca\,II\,H\&K.  In addition, Doppler shifts may
darken them further by shifting the line core into the passband.
A.\,G.~de Wijn has suggested that H$\alpha$ outer-wing darkness may also
result from excessive H$\alpha$ line width due to high temperature.

\begin{figure}
  \centering
  \includegraphics[width=\textwidth]{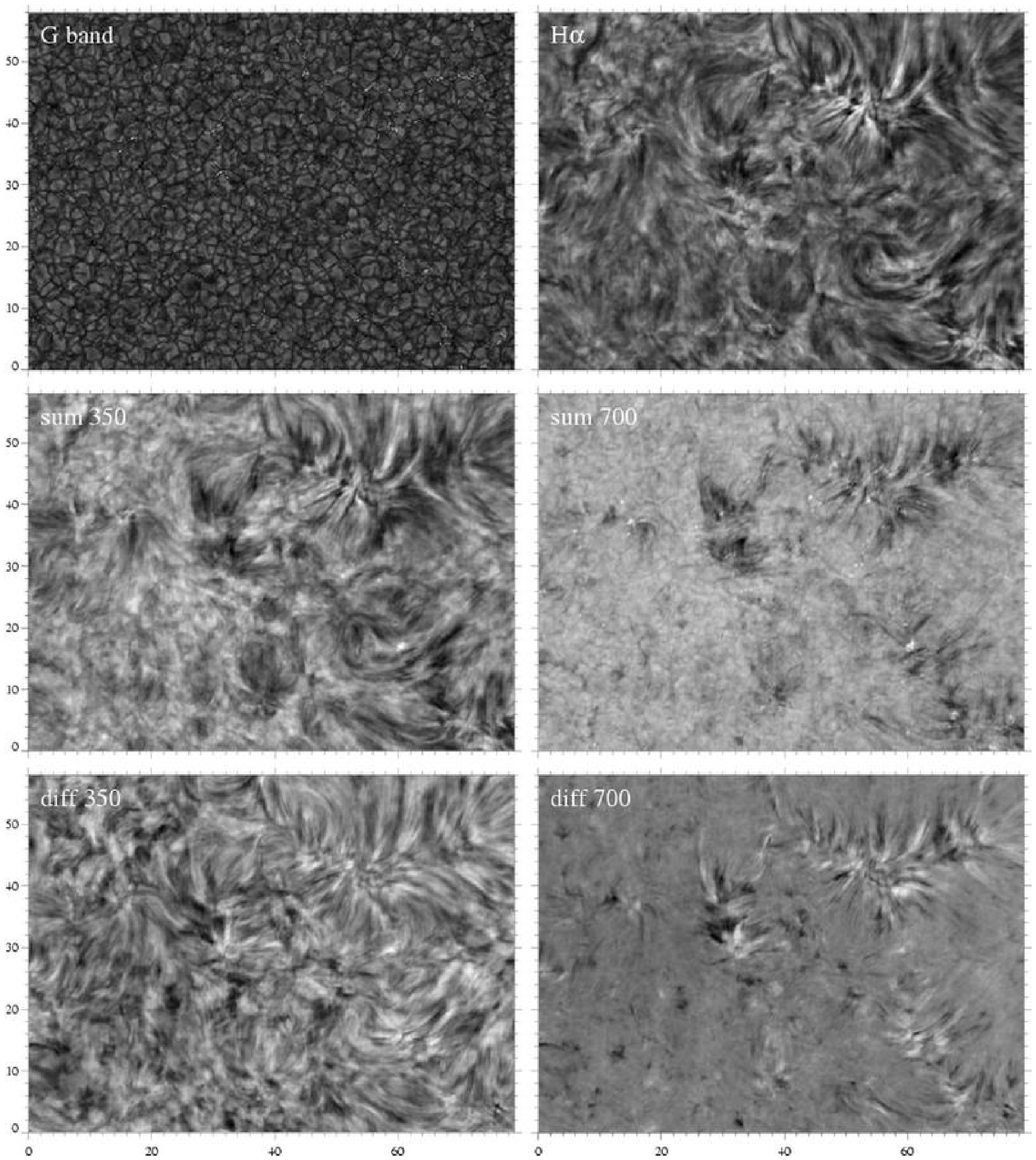}
  \caption[]{\label{rutten-fig:quiet}
  Very quiet disk-center scene taken with the DOT on October 19, 2005.
  {\em Upper row\/}: G band and H$\alpha$ center.  {\em Middle row\/}:
  sums of H$\alpha$ wing brightness at $\Delta \lambda = \pm 350$ and
  $\pm 700$~m\AA.  {\em Bottom row\/}: differences of these wing pairs
  (Dopplergrams, blueshift dark).  Scales in arcsec.  Courtesy
  P.~S\"utterlin.
}\end{figure}

Let us now move to disk center.  Figure~\ref{rutten-fig:network} shows
an area containing active network in Ca\,II\,H center and H$\alpha -
0.3$\,\AA.  The slender bright stalks jutting out from the network,
seen in Ca\,II\,H but only at high angular resolution, are the on-disk
representation of the near-limb straws.  Blinking shows that they
usually coincide with the lower ends of bright H$\alpha$ fibrils.
Many of the latter span much further out across the internetwork.  In
Ca\,II\,H the latter is dark with reversed granulation and therefore
photospheric or clapotispheric but not chromospheric.  H$\alpha$ shows
fibrils over much longer lengths than Ca\,II\,H, implying large heights
since Ca\,II\,H straws are preferentially upright.  However, in the
upper-right corner of this cut-out field the H$\alpha$ wing shows short
and grainy fine structure in a very quiet area, possibly as
chromospheric transparency into either the top of the clapotisphere or
the deep photosphere.

Figure~\ref{rutten-fig:quiet} display an extremely quiet region.  The
H$\alpha$ line-center scene shows much less large-scale organization
than for more active areas.  Most fibrils are short and strongly
curved.  In many places the H$\alpha$ chromosphere appears optically
thin.  The two summed-wing images (second row) confirm this: the
$\Delta \lambda = -700$~m\AA\ one shows only parts of fibrils close to
the scarce and sparse network.  The $\Delta\lambda=-350$~m\AA\
Dopplergram (third row) shows as much confusion as the line-center
image.  These samples come from a 71-minute multi-wavelength DOT
sequence with half-minute cadence (one minute for the H$\alpha$ wing
wavelengths between which the filter switched alternatively).  Movies
from these data show that this cadence is much too slow.  The quietest
areas on the solar surface are least stable in H$\alpha$.

\begin{figure}
  \centering
  \includegraphics[width=\textwidth]{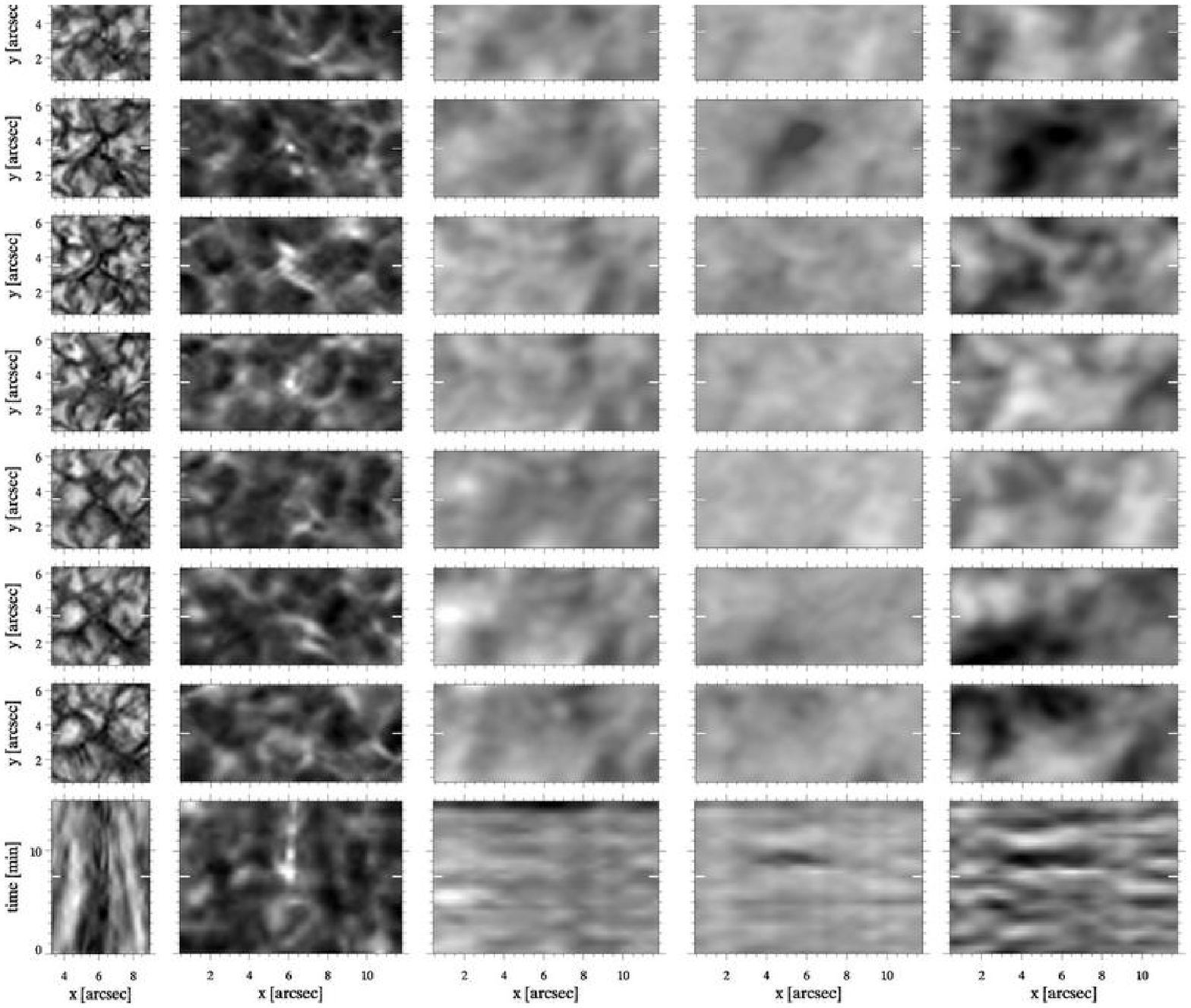}
  \caption[]{\label{rutten-fig:event} 
  H$\alpha$ response to an acoustic event, from DOT images taken on
  October 14, 2005.  The first seven rows are small image cutouts,
  respectively G band, Ca\,II\,H center, H$\alpha$ center, and H$\alpha$
  Dopplergrams from $\Delta\lambda = \pm 700$~m\AA\ wing pairs and
  from $\Delta\lambda = \pm 350$~m\AA\ wing pairs.  Bright implies
  blueshift in the latter two.
  The cutouts are wider for Ca\,II\,H and H$\alpha$ in order to show more
  context in the $x\!-\!t$ time slices in the bottom panels. The latter
  show the brightness evolution along the horizontal cut through the
  center of the subfield defined by the white markers in each image
  cutout.  The time step between consecutive image rows is one minute,
  with time increasing from bottom to top in correspondence with the
  time direction in the slices.  The image sequence is centered in time
  (fourth row, white markers in the slices) on the first appearance of
  a bright Ca\,II\,H grain.  Courtesy B.~van Veelen.
}\end{figure}

Figure~\ref{rutten-fig:event} indicates why.  The occurrence of an
``acoustic event'' is diagnosed by the sudden appearance of a bright
repetitive grain in Ca\,II\,H.  It follows on the squeezing away of a
small granular shard by converging large granules, just as the
``collapsars'' of
  \citet{rjr-2000ApJ...541..468S}. 
The H$\alpha$ line-center brightness does not react markedly but there
is large response in both Doppler samplings, most clearly seen in the
time slices: sudden onset of oscillation wave trains with upward
increasing amplitude.  Their spatial extent, 4~arcsec, is much wider
than the piston or the Ca\,II\,H grain, suggesting that an extended
piece of elastic canopy responds to the shock buffeting from below.
This area is also quiet, indeed without evident fibrilar structuring.
The time slices indicate that most of the small-scale H$\alpha$
brightness patterning is of oscillatory nature.

\begin{figure}
  \centering
  \includegraphics[width=65mm]{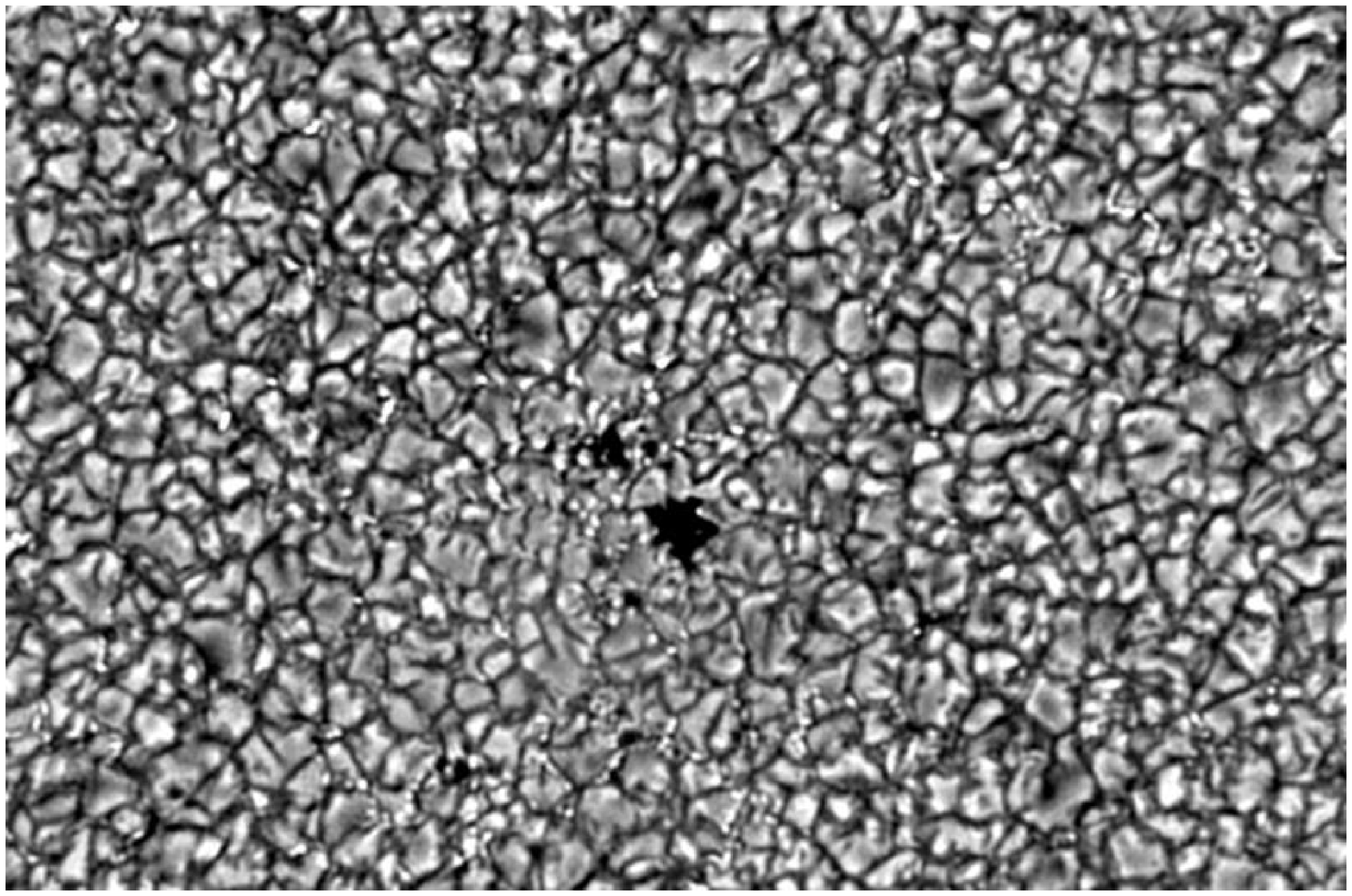}
  \hspace{0mm}
  \includegraphics[width=65mm]{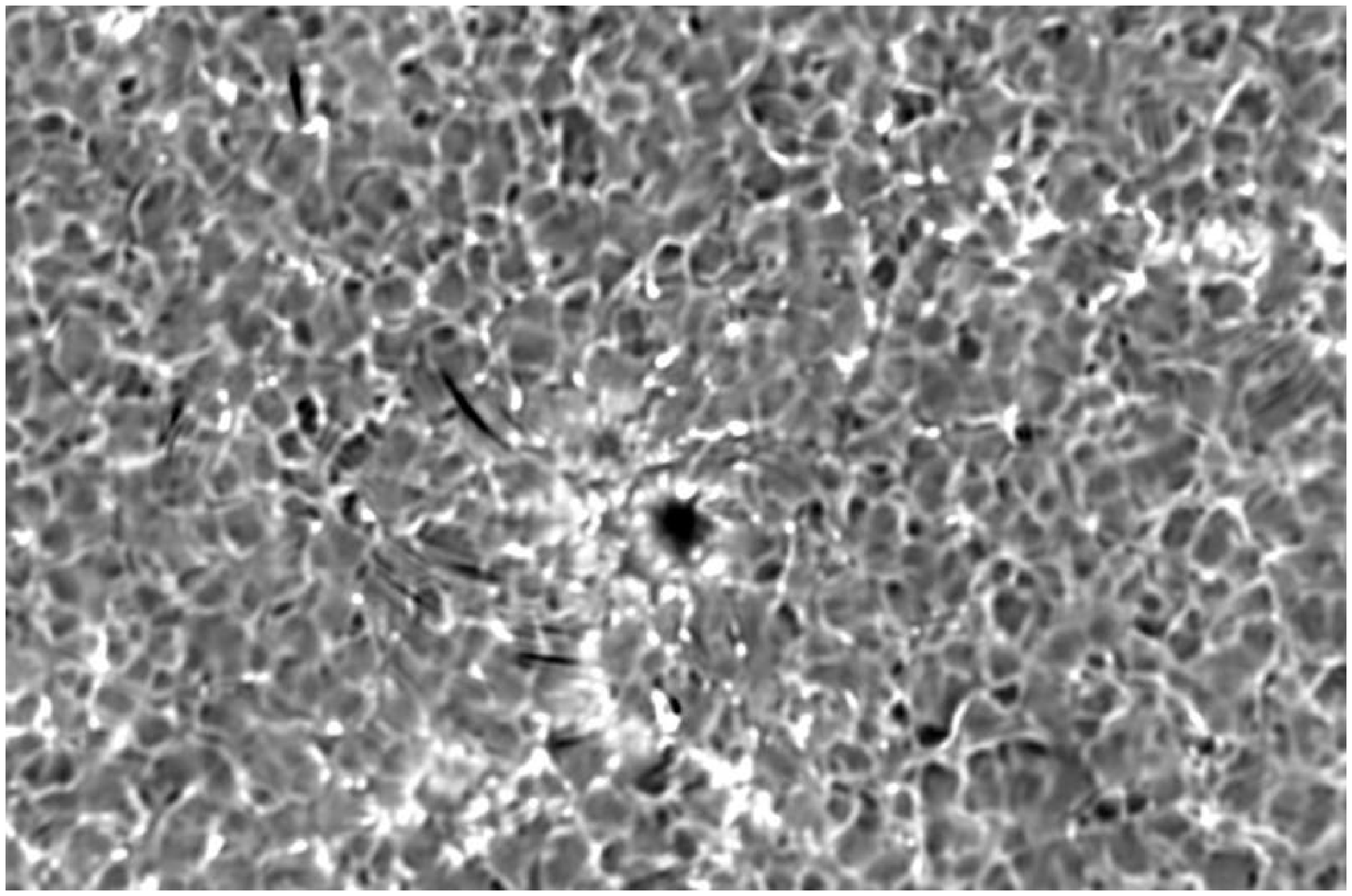}\\[2mm]
  \includegraphics[width=65mm]{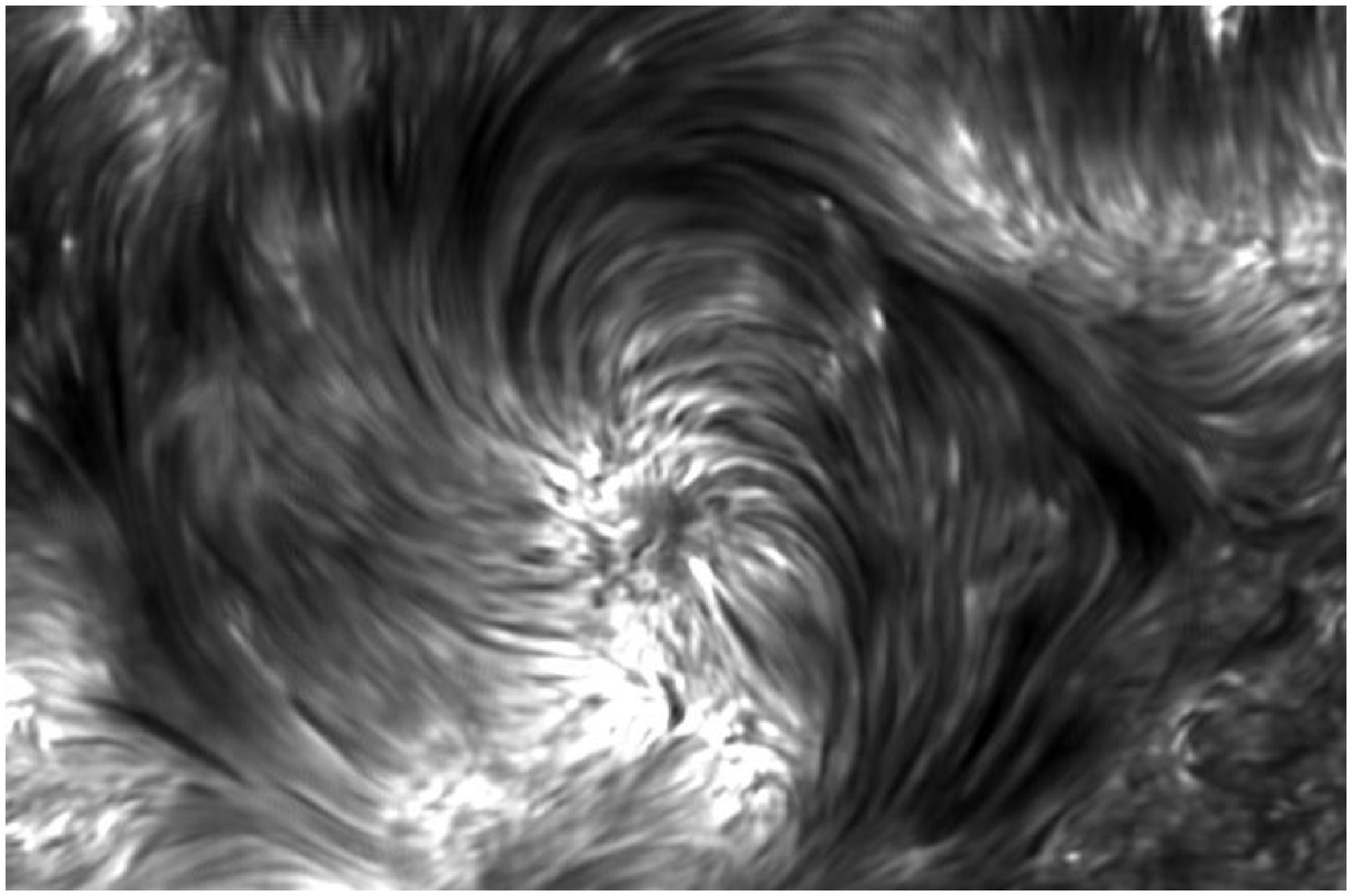}
  \hspace{0mm}
  \includegraphics[width=65mm]{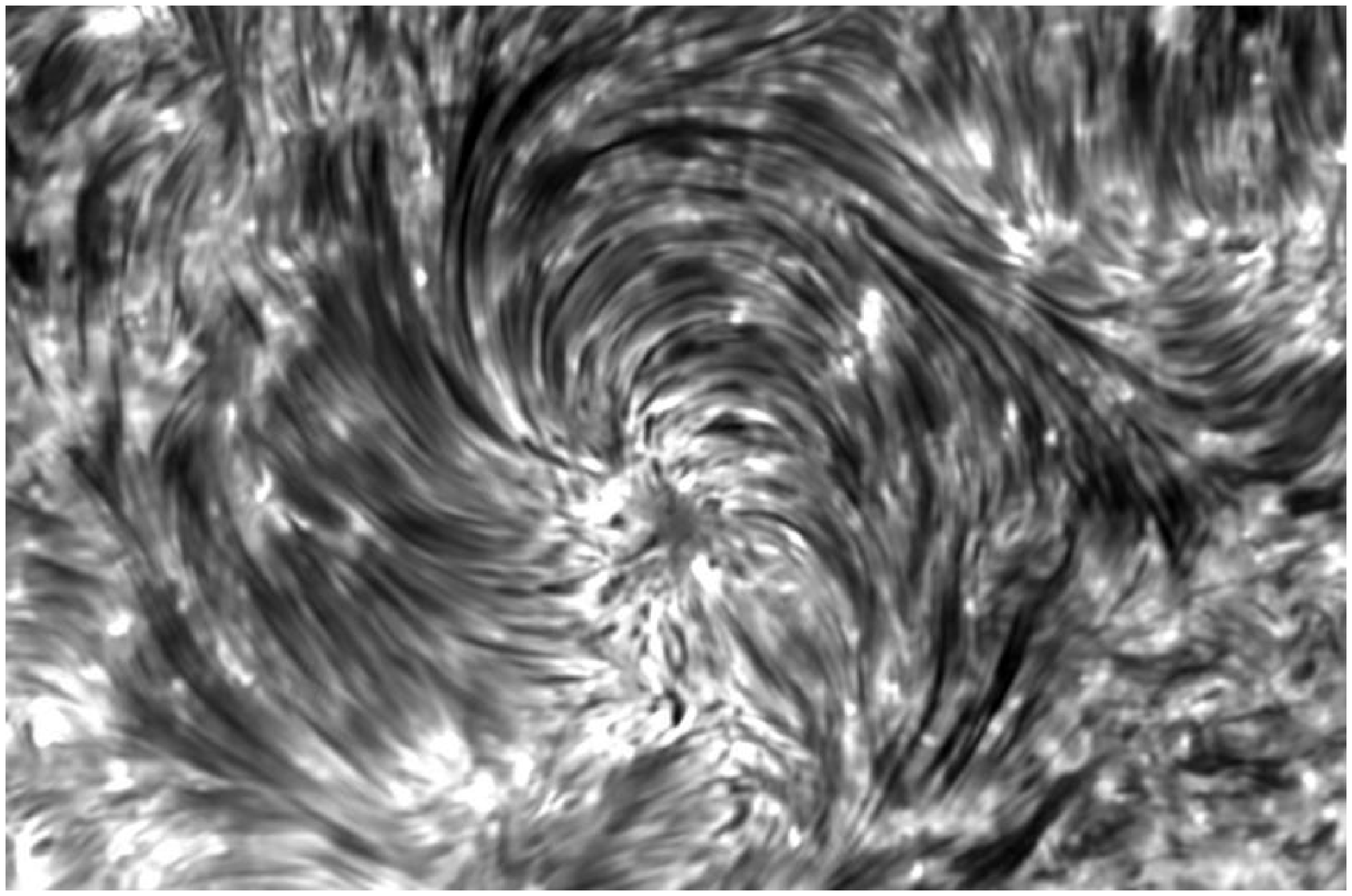}
  \caption[]{\label{rutten-fig:ibis}
  Near-simultaneous images taken with IBIS at the DST on October 1,
  2005 using adaptive optics and speckle reconstruction, clockwise in
  white light and in Ca\,II\,8542\,\AA\ at $\Delta \lambda =
  -600$, $-200$, and $0~$m\AA\ from line center.  Field about
  $60\times40$~arcsec$^2$.  Courtesy G.~Cauzzi.
}\end{figure}

\subsection{Ca\,II IR chromosphere}
Figure~\ref{rutten-fig:ibis} shows the chromosphere as it appears in
Ca\,II\,8542\,\AA.  The bottom panels resemble H$\alpha$ rather than
Ca\,II\,H in Fig.~\ref{rutten-fig:network}.  At various occasions I have
wondered how this subordinate calcium line can be more chromospheric
than its resonant H\&K siblings with larger opacity.  The
habitual answer of G.~Cauzzi, H.~Uitenbroek and M.~Carlsson is that
the sizable excitation energy of its lower level gives larger
temperature sensitivity and that the steep flanks of its narrower line
profile give larger Doppler sensitivity, and that these combine to
pick up fibril signatures that are less evident in H\&K filtergrams.
Reardon et al.\ illustrate the point further on
p.~151\,ff
in these proceedings.

In large active regions one can often trace dark fibrils in Ca\,II\,H if
one knows where they are from H$\alpha$, but they appear much clearer in
Ca\,II\,8542, with enhanced small-scale contrast in the inner-wing
panel of Fig.~\ref{rutten-fig:ibis}, presumably from Dopplershifts
(cf.\ Cauzzi et al.\ on 
p.~126\,ff
The
outer-wing panel shows the upper-photospheric mesh of reversed
granulation with acoustic brightenings
  (cf.\ \cite{rjr-Rutten2003e}) 
very markedly, suggesting that the temperature sensitivity does not
cause as wide a formation gap as the H$\alpha$ jump from normal (or
flattened, see
  \cite{rjr-2006A&A...460..301L}) 
granulation to chromospheric fibrils.

Combining H$\alpha$ and Ca\,II\,8542 with full profile sampling may
permit disentangling Dopplershift, thermal line broadening, and source
function variation since their atomic mass difference produces large
difference in thermal Dopplerwidth.  Dopplerwidth measurement may
provide a more direct handle on fibrilar temperature than fibril
brightness, most awkwardly set by NLTE opacity and source
function complexities for H$\alpha$.

\begin{figure}
  \centering
  \includegraphics[width=11cm]{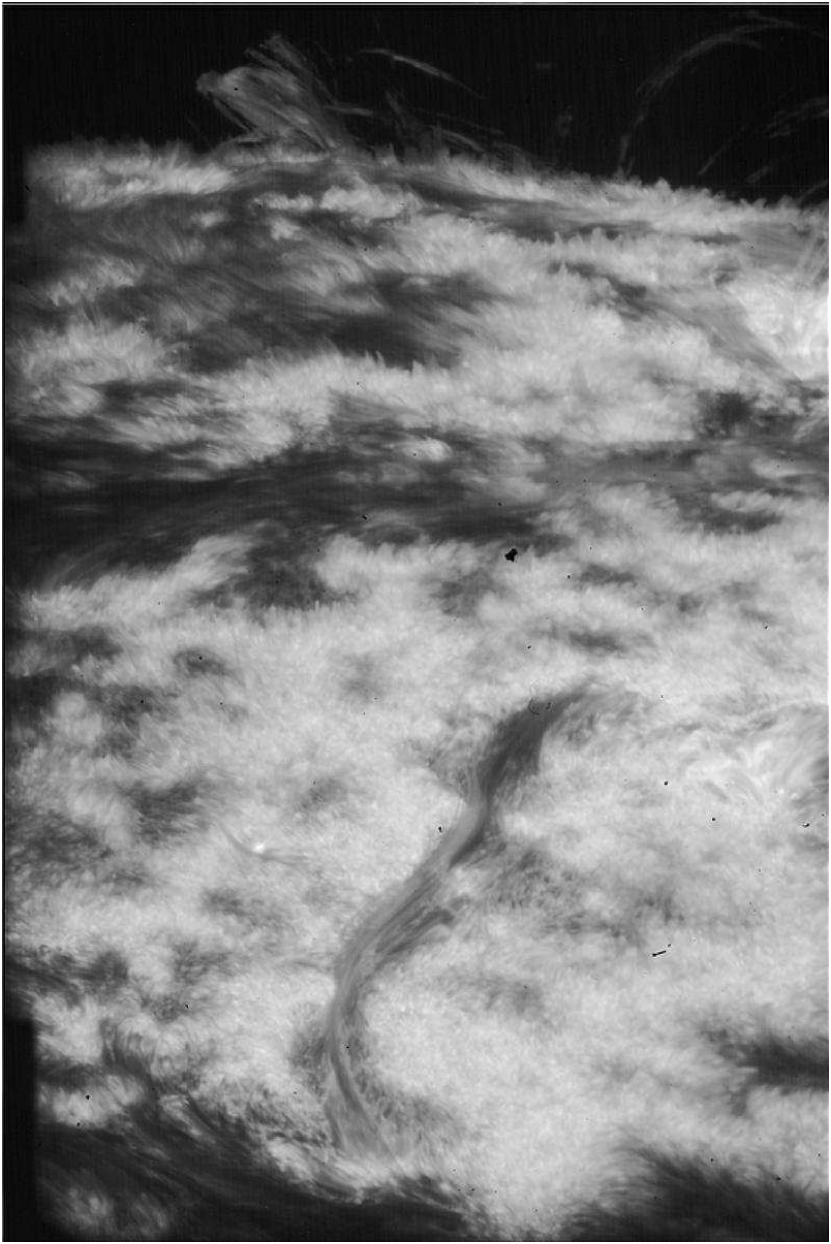}
  \caption[]{\label{rutten-fig:vault}
  Ly-$\alpha$ image from the second flight of the VAULT rocket 
  telescope
  (\cite{rjr-2001SoPh..200...63K}; 
   \cite{rjr-2001ApJ...563..374V}; 
   \url{http://wwwsolar.nrl.navy.mil/rockets/vault}).
  Field $246 \times 384$~arcsec$^2$.  The passband contains the
  full line.  Courtesy A.~Vourlidas.
}\end{figure}

\subsection{Ly-$\alpha$ chromosphere}

Figure~\ref{rutten-fig:vault} shows a beautiful Ly-$\alpha$ image.
Towards the limb it is remarkably similar to the outer-wing H$\alpha$
scene (Fig.~\ref{rutten-fig:limb}) but with reversed contrast: bright
hedge rows of short upright fibrils jut out at network borders of
cells; the latter are covered by flatter and darker extended-fibril
canopies.  The rows of stubby fibrils appear similar to
oscillation-loaded dynamic fibrils in H$\alpha$.  They seem optically
thick so that their brightness implies enhanced source function,
either through $S = (1-\varepsilon) J + \varepsilon B$
resonance-scattering with small $\varepsilon$ or through Balmer and
higher recombination adding an $\eta B^\ast$ term.  Both mechanisms suggest high temperature.  The
probable identity of H$\alpha$ and Ly-$\alpha$ fibrils implies that the
latter are thin hot sheaths around the former.

The extended active-region plage towards the bottom of the image 
appears very grainy, suggestive of mossy plage in TRACE 171\,\AA\
movies 
  (e.g., \cite{rjr-1999ApJ...519L..97B}; 
         \cite{rjr-1999ApJ...520L.135F}; 
         \cite{rjr-1999SoPh..190..419D}). 
%

\begin{figure}
  \centering
  \includegraphics[width=65mm]{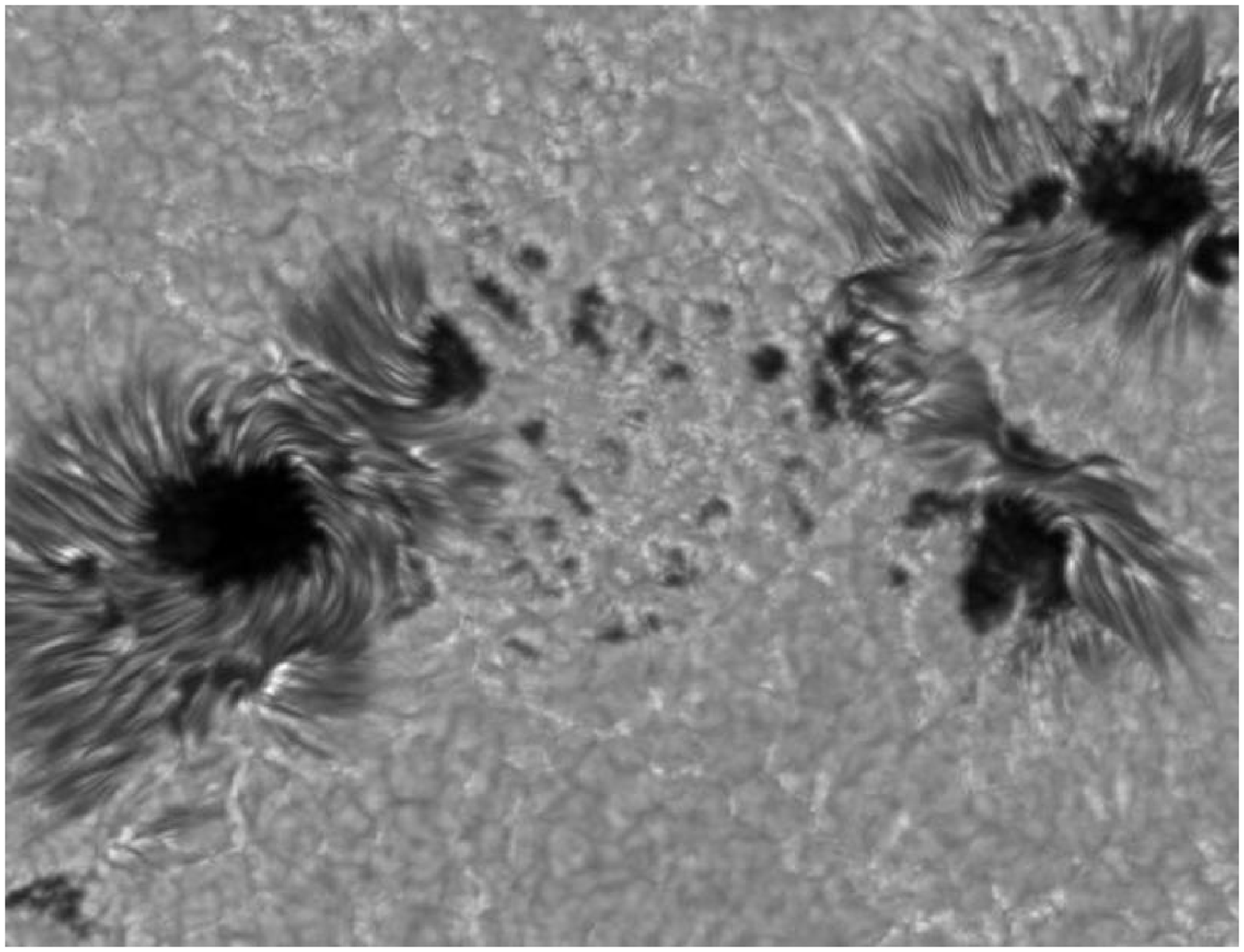}
  \hspace{0mm}
  \includegraphics[width=65mm]{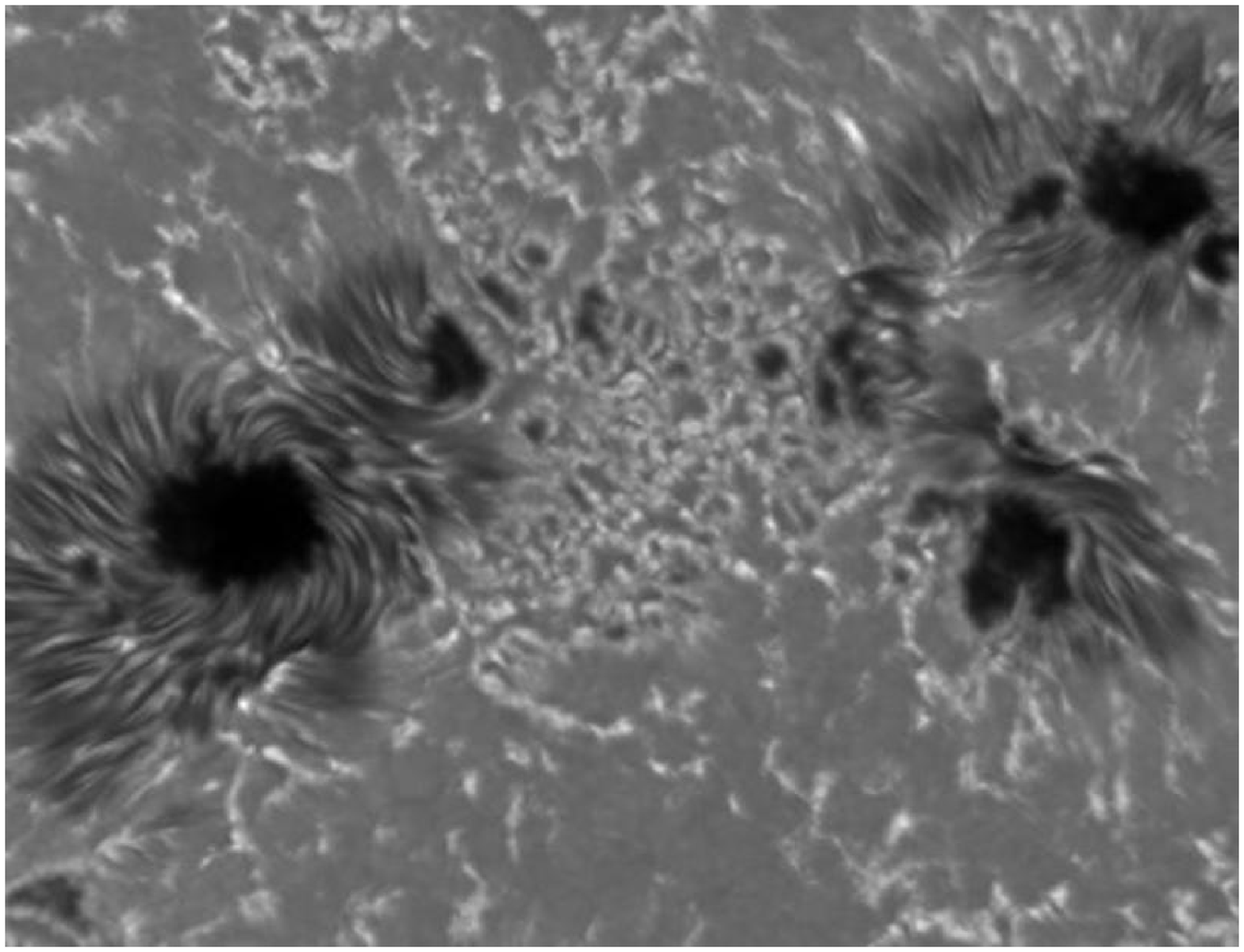}\\[1.5mm]
  \includegraphics[width=65mm]{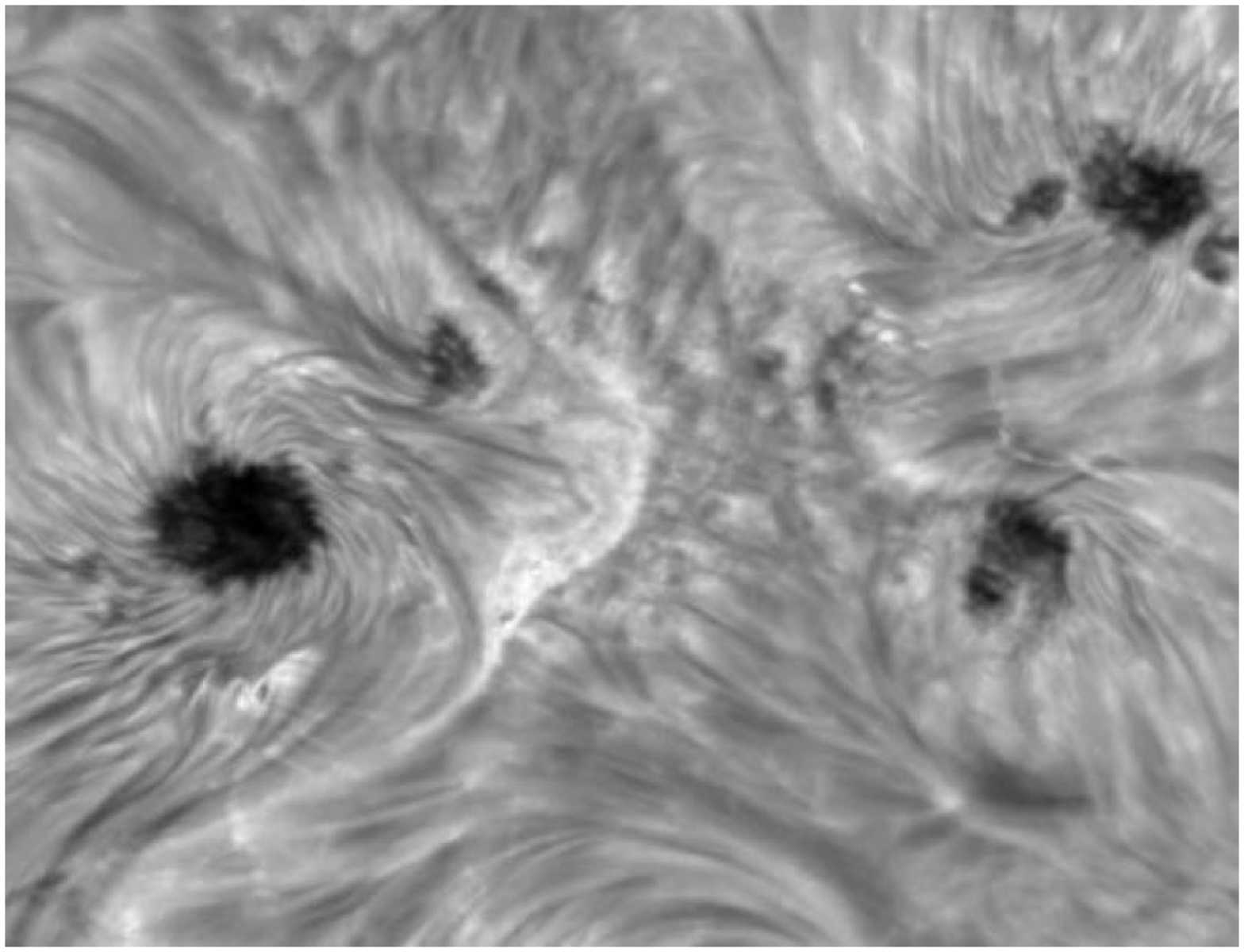}
  \hspace{0mm}
  \includegraphics[width=65mm]{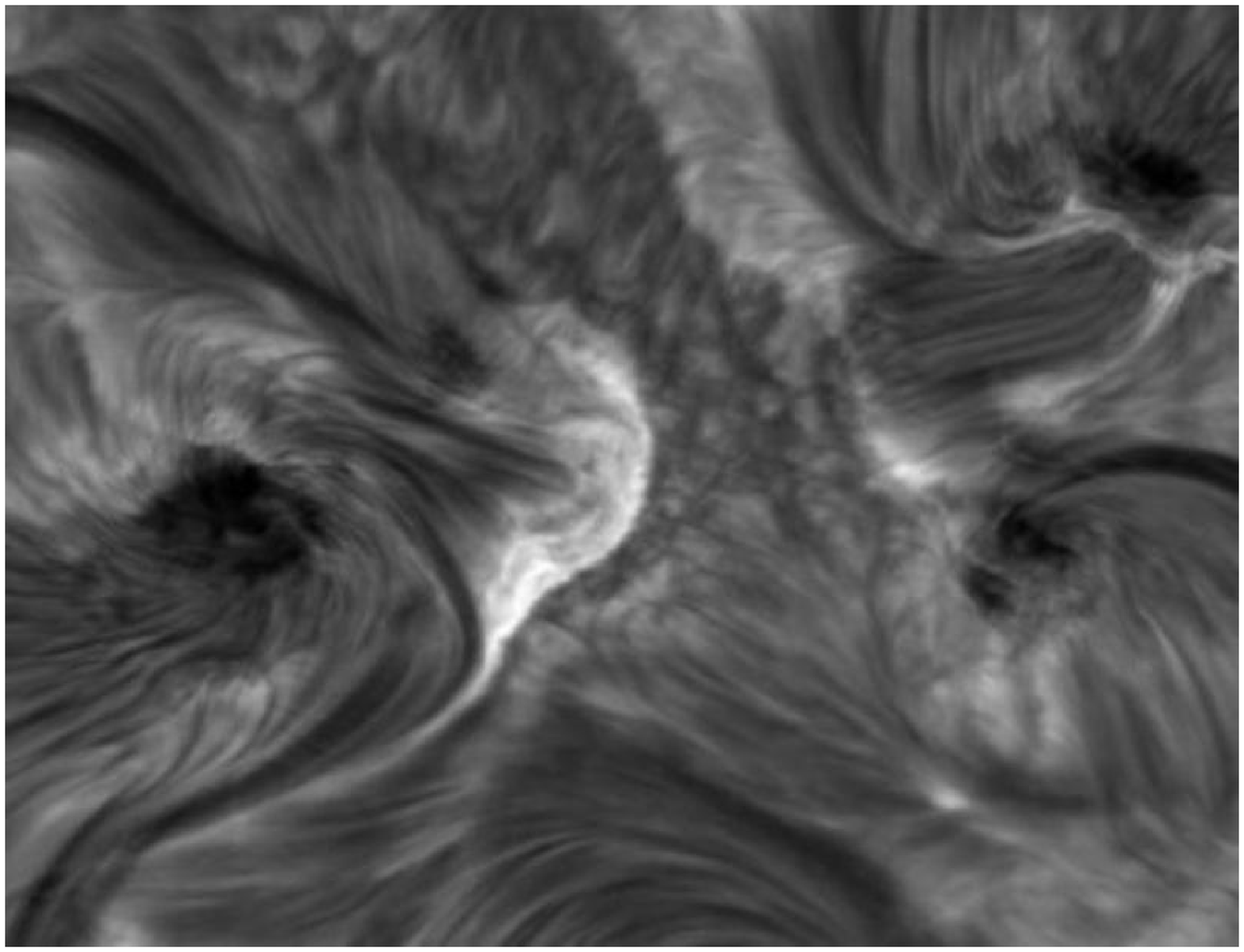}\\[1.5mm]
  \includegraphics[width=65mm]{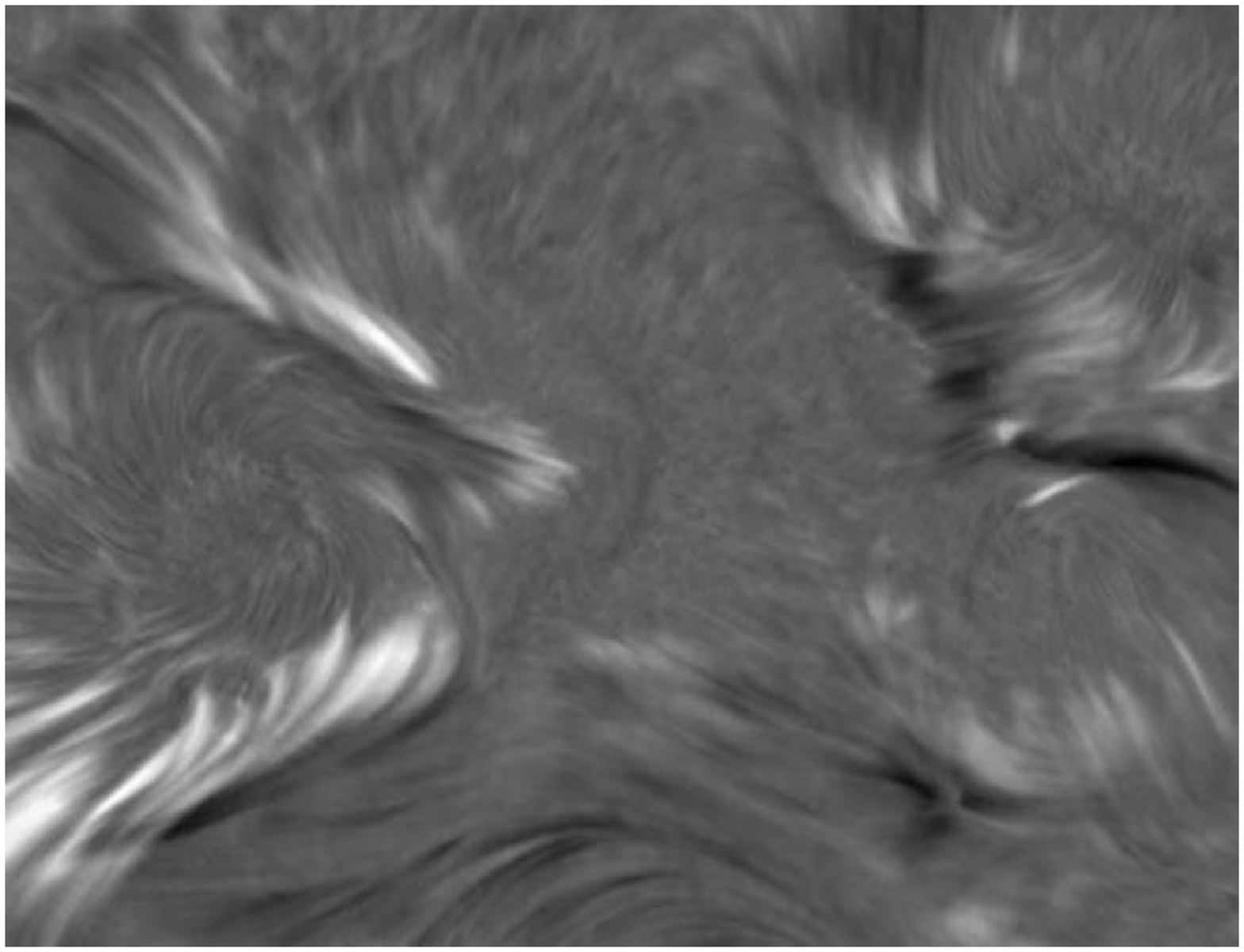}
  \hspace{0mm}
  \includegraphics[width=65mm]{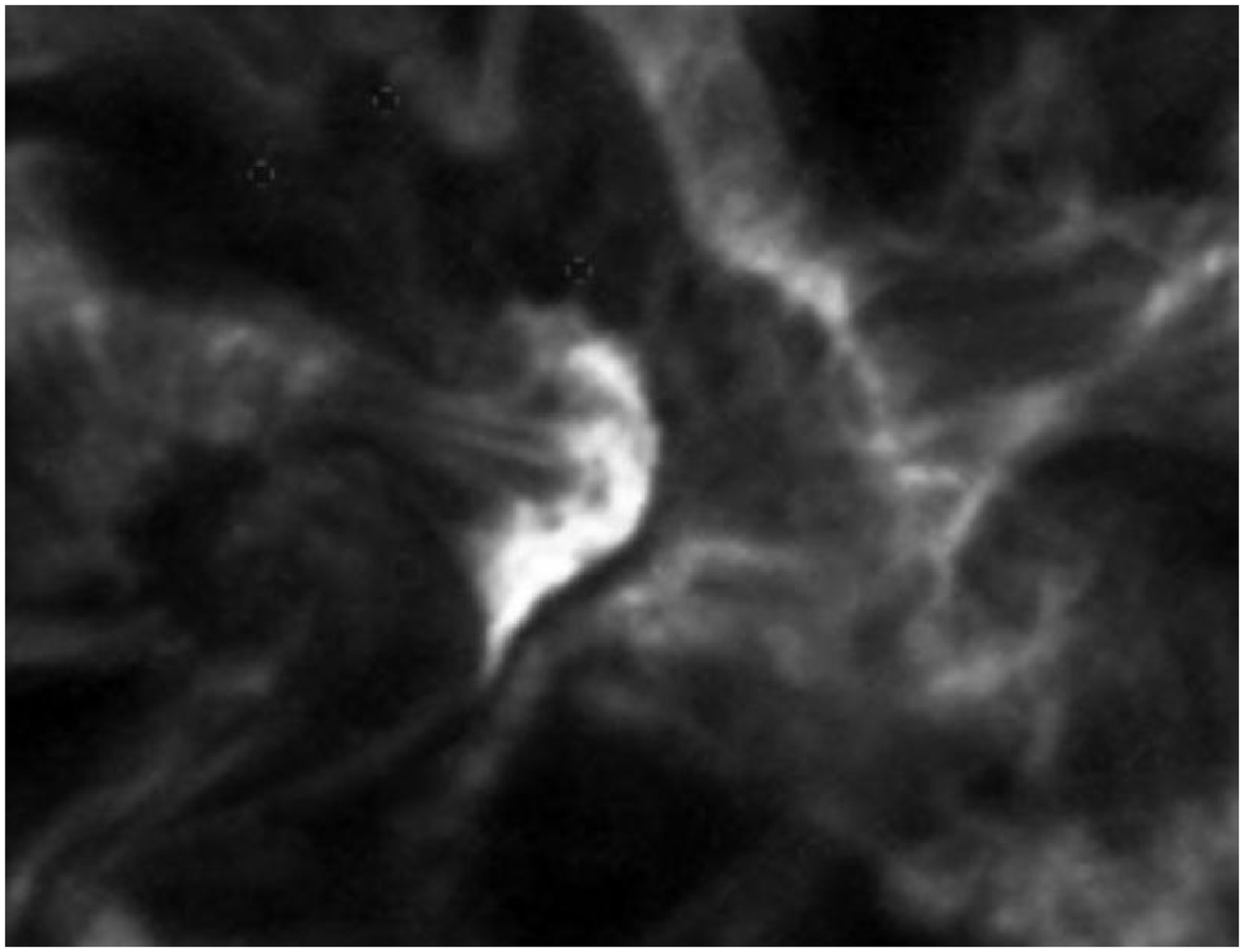}
  \caption[]{\label{rutten-fig:ha-171}
  One-hour temporal averages of co-aligned DOT and TRACE image
  sequences taken on July 9, 2005.  {\em Top\/}: G band; Ca\,II\,H at
  $\Delta \lambda = -2.35$\,\AA\ from line center.  {\em Middle\/}: sum
  of H$\alpha$ at $\Delta \lambda = \pm 500 $~m\AA; H$\alpha$ line
  center.  {\em Bottom\/}: difference of H$\alpha$ at $\Delta \lambda =
  \pm 500 $~m\AA\ (Dopplershift, upward dark); TRACE 171\,\AA.  Field
  size $78 \times 60$~arcsec$^2$.  The bright blob in H$\alpha$ and
  Fe\,171 is sharply delineated at right, presumably at the neutral
  line through this active region.  The H$\alpha$ wing difference
  indicates marked absence of large Dopplershifts in this area,
  without fibrilar structuring (best seen by zooming in with a pdf
  viewer).  Courtesy A.\,G.~de Wijn.
}\end{figure}

\subsection{H$\alpha$ and Fe\,171 chromosphere}

Finally, Fig.~\ref{rutten-fig:ha-171} compares the appearance of an
active region in H$\alpha$ images from the DOT and in Fe\,171\,\AA\
images from TRACE (movie: \url{http://dot.astro.uu.nl/movies}).  The
G-band image illustrates that granulation remains visible even in
one-hour averaging.  The Ca\,II\,H wing image represents an unsigned
magnetogram because reversed granulation vanishes better through
temporal averaging.  This was done here to emphasize the overall
bright-dark patterns in the remaining images.  They are strikingly
similar between H$\alpha$ line-center brightness and Fe\,171
brightness.  The first should sample $10^4$~K gas, the latter $10^6$~K
gas; the close similarity therefore needs consideration.

H$\alpha$ is bright either due to large chromospheric emissivity or due
to sufficiently small chromospheric opacity that one sees into the
underlying deep photosphere.  Note that the latter is certainly the
case in umbrae since they display umbral dots even at H$\alpha$ line
center (also in this figure), implying absence of chromospheric
opacity.  Bright-dark contrast between otherwise similar fibrils may
come from larger thickness of the darker ones, having lower scattering
source functions at their surface.  Excess emissivity may result from
excess excitation and from excess recombination.  The latter may arise
in steep-gradient neutral-to-coronal interfaces or through the Zanstra
mechanism as for planetary-nebulae Balmer emission.

Fe\,171 is bright solely through thermal photon creation but is
dark in two ways: either through absence of emissivity along the line
of sight into the black photospheric or sky background\footnote{The 
  term ``volume blocking'' is nonsensical since a volume doesn't block
  by itself.  It is used to express lack of emissivity along a line of
  sight through some volume, an ``emissivity void'' -- but a void
  doesn't block either; lack of emissivity cannot be expressed as
  opaqueness.},
or through blocking of thermal background brightness by foreground
bound-free out-of-the-passband scattering in the H\,I, He\,I and/or
He\,II continua as explained in Fig.~10 of
  \citet{rjr-Rutten1999d}. 

The bright blob at image center appears similarly in Fe\,171, at
H$\alpha$ line center, and in the H$\alpha$ wing summation.  The Fe\,171
blob suggests a relatively dense cloud of 10$^6$~K gas.  The H$\alpha$
brightness either results from chromospheric transparency or from
chromospheric emissivity.  Whichever of the two, it must correlate
with the presence of very hot gas.  Hydrogen ionization may explain
excess transparency, recombination excess emissivity.

\begin{figure}
  \centering
  \includegraphics[width=8cm]{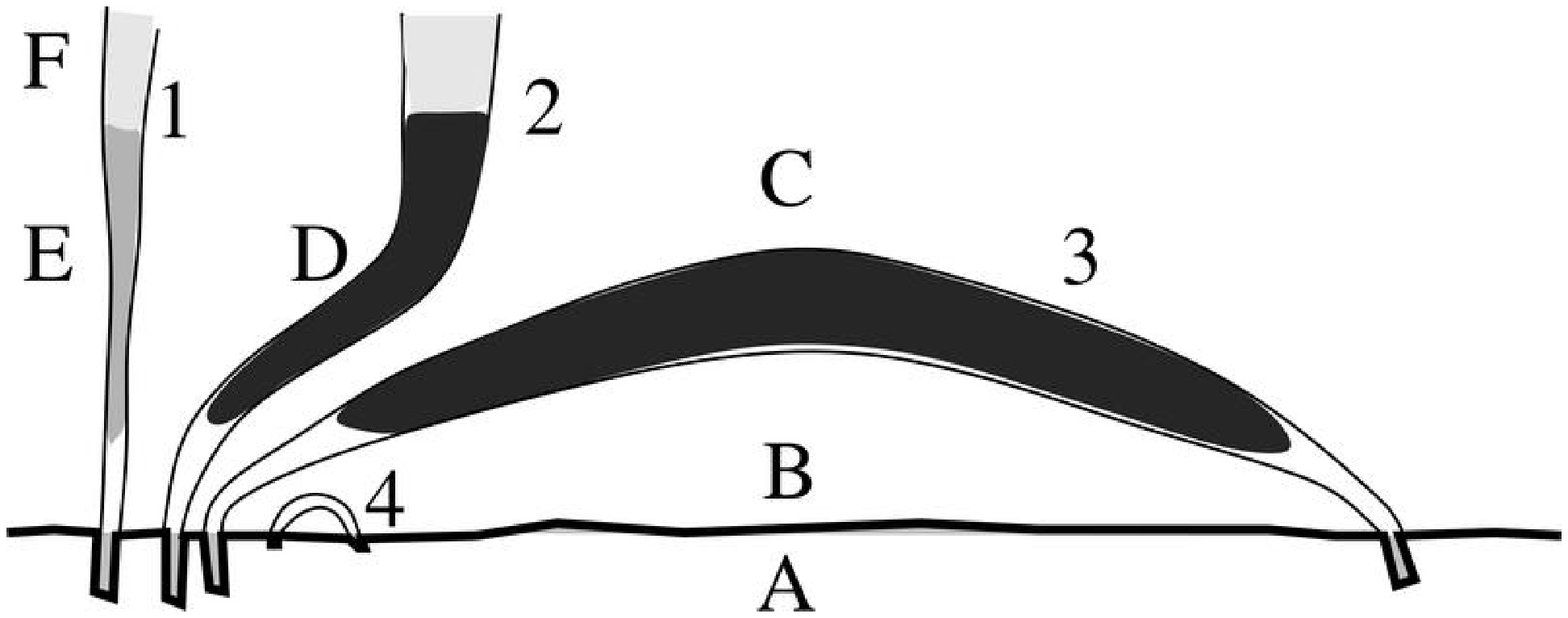}
  \caption[]{\label{rutten-fig:cartoon}
  Different types of fibrils together constituting the chromosphere.
  Rough temperatures: D$\sim$10$^4$\,K, E$\sim$10$^5$\,K,
  F$\sim$10$^6$\,K.  Outside regimes: A = photosphere with
  normal/reversed granulation and relatively empty magnetic elements,
  B = subcanopy ``clapotisphere'' pervaded by acoustic shocks
  but otherwise cool, C = coronally transparent in H\&K, H$\alpha$,
  and Ly-$\alpha$.
  Type~1: bright upright network straws
  opening into coronal plasma.  
  Type~2: dark H$\alpha$ fibrils bending upward into hot plasma
  from unipolar crowding.  
  Type~3: dark H$\alpha$ fibrils spanning across cell interiors
  in bipolar network outling magnetic canopies.
  Type~4: short weak-field near-network loops postulated by
  \citet{rjr-2003ApJ...597L.165S}. 
  From
  \citet{rjr-Rutten2007a}. 
}\end{figure}

\section{Speculations}

My impressions from the images in the previous section are:
\leftmargini=3ex \begin{itemize} \vspace{-1ex} \itemsep=-0.3ex

  \item the chromosphere consists of the mass (or mess) of fibrils
        observed in H$\alpha$;

  \item cell-spanning fibrils outline magnetic canopies.  Above quiet
        cell centers they are jostled and kicked up by
        upward-propagating shocks, producing an extended cool
        clapotisphere underneath;

  \item slanted fields jutting out from network and plage suffer
        oscillatory cool-gas loading to become dynamic fibrils
       (\cite{rjr-2007ApJ...655..624D}) 
        and EUV shutters
        (\cite{rjr-2007ApJ...654.1128D}). 
        These possess transition-interface sheaths that are optically
        thick in Ly-$\alpha$;

  \item more straight-up fields above network and plage show up as
        bright Ca\,II straws and bright H$\alpha$ fibril feet, likely
        through recombination emissivity.  Near the limb they appear
        as hedge rows that are bright in Ca\,II and Ly-$\alpha$ but dark
        in the outer H$\alpha$ wings from bright background, large
        Dopplershift, and/or large thermal broadening.

  \item near-vertical fields in active regions produce graininess in
        Ly-$\alpha$, mossy plage, and brightness correspondence between
        H$\alpha$ and Fe\,171.  They provide locations where hot gas
        comes very close to the photosphere, presumably through
        downward conduction and lack of upward kicking.

\end{itemize}
These impressions are cartoonized in Fig.~\ref{rutten-fig:cartoon},
copied from
  \citet{rjr-Rutten2007a} 
where I summarized them without showing so many tell-tale images.

\section{Suggestions}

We obviously need improved H$\alpha$ imaging.  Combination with H$\beta$
and one or more of the Ca\,II IR lines is desirable.  Yet better is
multi-line chromospheric imaging with synchronous co-spatial
Dopplergrams and magnetograms of the underlying photosphere and with
TRACE/SDO EUV imaging of the higher-temperature scenes.  All this in
long-duration movie sequences.  Unfortunately, Hinode's tunable filter
seems not fulfilling its intended H$\alpha$ capability (but would
anyhow deliver only slow cadence due to telemetry limitations).

The H$\alpha$ images in
Figs.~\ref{rutten-fig:mosaic}--\ref{rutten-fig:event} illustrate the
necessity of high angular resolution.  Because H$\alpha$ fibrils appear
as high-lying Schuster-Schwarzschild clouds which may be optically
thin, there is no smallest-scale limit set by mean-free photon paths
or scattering lengths.  In addition, their dynamical behavior
necessitates taking sustained-quality image sequences at fast cadence.
The DOT does so admirably, but the
twice larger SST yields not only higher resolution but also better
signal-to-noise through better cameras, better alignment through
MOMFBD restoration
 (\cite{rjr-2005SoPh..228..191V}), 
and especially faster cadence because MOMFBD requires fewer frames
than speckle reconstruction.  
  \citet{rjr-2006ApJ...648L..67V} 
demonstrated that imaging cadence as fast as 1~fps is needed for some
H$\alpha$ dynamics, well beyond the traditional estimate of soundspeed
travel across a resolution element.

In addition, one should add line-profile sampling to disentangle the
complex cross-talk between opacity, source function, and Dopplershift
variations.  Chromospheric imaging at high resolution so becomes just
as photon-starved as photospheric spectropolarimetry at high
resolution, making fast-cadence profile-sampling narrow-band imaging a
second motivation for telescope aperture beyond the angular resolution
in reach of AO and post-detection processing.  Post-focus light
handling is presently done best by rapid-scan Fabry-P\'erot imaging.
Fiber field reformatting may enable 2D MOMFBD spectrometry in the
future
  (\cite{rjr-Rutten1999c}). 

The obvious desire to add ultraviolet spectrometry at high angular
resolution to sample hot fibril sheaths is presently unanswered:
no HRTS- or SUMER-like spectrometer is available for regular
co-pointing.  Even better would be integral-field ultraviolet profile
sampling because slits, whether scanning or sitting-and-staring, tend
to be at the wrong place at the wrong time.  The lack of such
ultraviolet instrumentation is the major longer-term deficiency in
chromospheric observing, making H$\alpha$ the key diagnostic by
default.

The simple question ``where is this line formed?'' discussed in
Section~\ref{rutten-sec:lineformation} is a na\"{\i}ve way of asking
``how well does my interpretation fit reality?'', a question well
beyond observation.  So let me end by addressing modeling options.

The SIR technique thanks its large success in its main application,
photospheric Stokes profile inversion, to the same reasons why
Holweger's line fitting made HOLMUL such a success in abundance
determination: LTE and smooth radial behavior are good assumptions for
photospheric iron lines.  Extension to scattering and sinusoidal waves
around $h\!=\!500$~km is doable (outside sunspots) within such
inversion approaches, but already the shocked clapotisphere below the
canopy needs forward modeling based on numerical simulations.

The fibrilar chromosphere surely requires simulation physics to
diagnose its structural and dynamical physics; the breakthrough
example is Hansteen's simulation in
  \citet{rjr-2007ApJ...655..624D}. 
The chromosphere cannot be treated by inversion before its physics is
understood, and then only if all fibrils are so similar in geometry,
temperature, density, etc., that multi-cloud modeling (with some sort
of NLTE radiative transfer in and between each) remains realistically
limited in parameter space.  The interpretational path to follow is
forward modeling based on 2D and 3D simulations.  However,
PANDORA-style 1D modeling suits to study radiative transfer in and
between fibrils with transition sheaths in full detail for working out
and testing simpler recipes for multi-D codes.  This
step is similar to studying transition-sheath radiation in and from
prominences as by
  \citet{rjr-2006A&A...459..651S}. 

Shortcutting through untested tractability assumptions is 
dangerous even though it may produce intrinsically interesting
``code-as-a-star'' papers.\\

\acknowledgements I thank the CSPM organizers for a very good meeting
and for inviting me to both speakership and editorship.  My
participation was funded by the Leids Kerkhoven-Bosscha Fonds and the
EC-RTN European Solar Magnetism Network.


\end{document}